\documentstyle[fleqn,epsfig]{article}

\begin{document}
\title{
\rightline{\small HUB--EP--96/59}
\rightline{\small revised}
Topology at the Deconfinement Transition\\
       Uncovered by Inverse Blocking  \\
       in $SU(2)$ Pure Gauge Theory \\
       with Fixed Point Action\thanks{Supported by
       a Visitorship of M.~F. and S.~T. at the
       Gra\-du\-ier\-ten\-kol\-leg
       ''Struk\-tur\-un\-ter\-su\-chun\-gen,
       Pr\"a\-zi\-sions\-tests und
       Er\-wei\-terun\-gen des Stan\-dard\-mo\-dells der
       Ele\-men\-tar\-teil\-chen\-physik''}
      }
\author{M.~Feurstein,
E.--M.~Ilgenfritz,
\thanks{Supported by the Deutsche Forschungsgemeinschaft under grant  Mu932/1-4}
  \\
M.~M\"uller--Preussker
and S.~Thurner \\
{\it Institut f\"ur Physik, Humboldt--Universit\"at zu Berlin, Germany}
       }

\maketitle
\begin{abstract}
Renormalization group transformations as discussed recently
in deriving fixed point actions are used to analyse the vacuum structure
near to the deconfinement temperature. Monte Carlo configurations
are generated using the fixed point action. We compare equilibrium
configurations with configurations obtained by inverse blocking from a
coarser lattice.
The absence of short range vacuum fluctuations in the latter
does not influence the string tension.
For the inversely blocked configurations we find the following:
({\it i}) the topological susceptibility
$\chi_{top}$
is consistent with the phenomenological value
in the confinement phase,
({\it ii})
$\chi_{top}$ drops across the deconfinement transition,
({\it iii}) density and size of instantons are estimated,
({\it iv}) the topological
density is found to be
correlated to
Abelian monopole currents and ({\it v}) the
density of spacelike
monopole currents becomes a confinement order parameter.
\end{abstract}

\section{Introduction}

Not much is known from first principles about the topological
structure at the thermal phase transition in Quantum Chromodynamics. 
How 
this structure changes at the deconfinement phase transition is not
even clear in pure gauge theory. What we mean by ''topological
structure'' is the 
characterization of Euclidean field trajectories in terms of instantons 
\cite{CDG,Shuryak} and 
Abelian monopoles \cite{Mandelstam,tHooft} and their interrelation. 
Changes of this structure are
expected
to be the common mechanism for deconfinement and restoration of chiral
symmetry. Understanding this mechanism is important for
understanding the QCD vacuum.

There was a
wave of attention recently for improving
lattice actions for gauge fields \cite{Improved} (and fermions). 
Part of the interest was motivated 
to minimize finite lattice spacing corrections in simulation results
obtained on coarse lattices.
One particular approach towards improvement 
was guided by the
desire to construct a perfect action being a 
fixed point action \cite{FixedPointG,FixedPointF}
with
respect to some real space renormalization group (RG)
transformations. In this context the concept of inverse blocking
came up which is the essence of a new method,
recently proposed by DeGrand et al. \cite{DeGrand96.1,DeGrand96.2}.
In particular it improves the determination of the
topological charge of a lattice
configuration.

In the present paper we want to explore 
the capability of this method, 
to resolve more details and different aspects
of topological structure.
In the neighbourhood
of the deconfinement transition of the
pure $SU(2)$ gauge theory,
we want to study the space-time distribution of
topological charge, correlations of the density with itself and with
Abelian monopole currents.

The topological charge 
density, when defined for quantum fields,
is obscured by ultraviolet fluctuations which must
be removed in some way. Cooling \cite{Cooling} has been invented
for this purpose long ago. 
According to the
Abelian dominance scenario at large distances \cite{Mandelstam,tHooft}
Abelian magnetic monopoles
are closely related to confinement. 
The density \cite{Bornyakov1} and anisotropy \cite{Bornyakov2}
of these magnetic currents are studied
in the maximally Abelian gauge \cite{Kronfeld}. 
These currents are
quantized by construction \cite{DeGrandToussaint}. 
This does not exclude the possibility that some part of these
currents are lattice artefacts which do not correspond to long distance 
physics. 
Correlations indicating a close relation between topological charge
and Abelian magnetic monopole currents have been studied,
not addressing this question.
They have been studied so far for
particular instanton configurations \cite{Bornyakov3} and,
as far as the quantized vacuum is concerned, the cooling method has been 
employed \cite{Vienna}. However, cooling is known to 
change the topological structure and has to be used with care.

A closer examination shows that 
the method of Refs. \cite{DeGrand96.1,DeGrand96.2}
gives not only the topological charge but
also permits a
reasonable definition 
of the topological density of generic,
equilibrium lattice fields  
with minimal effort. 
A better 
resolution of topological and monopole structure is possible 
without recourse to cooling.
While the method 
is applied  
to extract the topological structure, 
the monopole activity is changed, too, roughly in proportion
to the topological activity (to be defined below). We shall see 
that part of it becomes
an order parameter of the confinement/deconfinement transition, 
according to the alleged role
monopoles play for confinement.  

According to the scope of our present study,
let us
sketch the background of this
approach \cite{FixedPointG} restricting the 
discussion to the case of pure gauge field
theories. The starting point is a block spin transformation
(here of scale factor two) which maps a fine lattice configuration of
link matrices $U_{x,\mu}$ to a coarse lattice configuration
with link matrices $V_{x,\mu}$
(each single link  being an element of the gauge group $SU(N)$).
The second ingredient is a general type of action $S_{FP}$,
written on both lattices 
in terms of Wilson loops (evaluated in various representations).
The classical perfectness of this action is guaranteed if
there exists an inverse blocking transformation  
$~V \rightarrow U~$
with respect to which the
action saturates the inequality
\begin{equation}\label{eq:inequality1}
~~~~~~~~   S_{FP}(U) + \kappa ~ T(U,V) \geq S_{FP}(V) ~~.
\end{equation}
Here, $T(U,V)$ is some non-negative functional related to the blockspin
transformation and $\kappa$ is a kind of Lagrange
multiplier. 
Once the action and other parameters have been determined, this
inequality leads to the target configuration
$~U$ of the inverse blocking by minimization of the l. h. s., 
the fixed point action with respect to $~U$, 
with a constraint 
derived from the coarse configuration $~V$.
Thus, $~U$ is found as the result of an
iterative relaxation process and cannot
be explicitely expressed.
We call the mapping $~V \rightarrow U$ one step of inverse
blocking. Similar to blocking, inverse blocking can be iterated. This
leads to a sequence of 
interpolating configurations on finer and finer lattices.

Without giving details, we mention 
that the saturated inequality (\ref{eq:inequality1}) 
is constructive for finding the perfect action itself. It has been 
used \cite{DeGrand96.1}
to optimize the ratios between the couplings in front of various loops in
different representations 
within the fixed point action $S_{FP}$, as well as
the other constants related to the block spin transformation. 

After this machinery has been set up, the procedure of inverse blocking
remains a useful tool of analysis.
Doing a sequence of inverse blocking steps one would be
able to relate
a continuum configuration to any given lattice configuration. 
The topological charge of this continuum configuration is {\it the
topological charge} related to the given lattice configuration
(the {\it classically perfect topological charge}). This 
prescription is an important conceptual step forward 
compared with previous techniques of measuring topological charge
on the lattice, eventually used in combination with 
cooling. 

Inverse blocking has been used before \cite{DeGrand96.1,DeGrand96.2}
only to measure the topological charge
of lattice configurations, by evaluating the 
Phillips--Stone (PS) charge \cite{PhillipsStone}
just on the
next finer level. The success indicates
that one step of inverse blocking can be  
sufficient to have access to the 
classically perfect topological charge
corresponding to the {\it continuum}
by means of a {\it lattice} definition.
Other lattice 
algorithms besides PS 
could also have been used, for instance 
in order to test the uniqueness of charge. 
We will use in this paper a geometric charge due to
L\"uscher \cite{Luescher,LuescherImplem} and the old
plaquette based (field theoretic or naive)
definition \cite{DiVecchia,MMPMakh}. Both have the advantage
to localize the
topological density in space-time which makes them suitable
for our present purpose.
 
The first concern in Ref. \cite{DeGrand96.1} 
has been that action and topological charge of artificially constructed
lattice instantons should be independent of size, if the
perfect action is used and the topological charge is measured as
described. This was demonstrated for instanton configurations
with $\rho > 0.8 \cdot a$.
There were no dislocations, with action 
$S/S_{I} < 6/11$ (entropic bound, instanton action $S_{I}=2 \pi^2$) and
non--zero topological
charge, among them.
However, the perfect action and the new
definition of
topological charge
are useful also at finite $\beta$.
In the second paper \cite{DeGrand96.2}
a (truncated) fixed point action
has been successfully tested for scaling of the deconfinement temperature,
of the torelon mass, of the string tension and of the topological
susceptibility. For the latter, it has been stressed that
inverse blocking has been indispensable to achieve 
reasonable
scaling.
This has nourished the hope that the
fixed point action and the inverse blocking prescription
for the topological charge can avoid
dislocations which otherwise (for instance in the case of Wilson's action)
are known to spoil the continuum limit of the topological 
susceptibility \cite{NoScaling}. 

Even the temperature dependence of the topological
susceptibility has remained controversial until recently. 
Armed with the new method, the  
topological susceptibility has been measured across the deconfinement 
transition \cite{DeGrand96.2} without the ambiguities of the 
cooling method \cite{DiGiacomoT1,IlgLATT94}. Inverse blocking is not
the only technique to avoid cooling. We only want to mention the
use of an improved lattice operator 
of the topological density \cite{ImprovedQ}. This approach
has given a much more
discontinuous drop \cite{DiGiacomoT2} at the deconfinement transition
than the inverse blocking method \cite{DeGrand96.2}.

Our paper is organized as follows. In section 2 we give details on the
fixed point action, on blocking
and inverse blocking.
In section 3 we present results on the string tension. For symmetric
lattices this serves to calibrate the lattice step size $a(\beta)$
in accordance to simulations using
the fixed point action. The finite temperature string tension is computed
(in the confined phase) from Polyakov line correlators.
These results are indicating that the string tension
is largely conserved by inverse blocking.
Section 4 contains our results on
the gross topological features. 
The topological susceptibility obtained on
inversely blocked configurations in the confined phase is 
quantitatively reasonable.
The temperature dependence of the topological susceptibility across
the phase transition is studied. Analysing inversely blocked configurations
we observe,
apart from the decay of the topological susceptibility,
the complete suppression of spatial Abelian monopole currents
in the deconfined phase.
In section 5 we present results on topological charge
density--density correlations and on
monopole--instanton correlations. We visualize 
clusters of topological charge together with their accompanying
monopole currents for
typical confining configurations. In closing, directions for further work
are pointed out in section 6.

\section{Fixed Point Action and Inverse Blocking}

\subsection{Action and Blocking}

The simplified fixed point action \cite{DeGrand96.2}
is parametrized in terms of only two
types of Wilson loops,
plaquettes $U_{C_{1}}=U_{x,\mu,\nu}$ (type $C_{1}$)
and tilted $3$-dimensional $6$-link loops
(type $C_{2}$) of the form
\begin{equation}\label{sixlinks}
U_{C_{2}}=U_{x,\mu,\nu,\lambda}=
U_{x,\mu}
U_{x+\hat{\mu},\nu}
U_{x+\hat{\mu}+\hat{\nu},\lambda}
U_{x+\hat{\nu}+\hat{\lambda},\mu}^+
U_{x+\hat{\lambda},\nu}^+
U_{x,\lambda}^+  ~~.
\end{equation}
It contains several powers of the corresponding linear action terms
as follows
\begin{equation}\label{eq:fp_action}
S_{FP}(U)=\sum_{type~ i} \sum_{C_{i}} \sum_{j=1}^{4}
w(i,j) (1 - {1 \over 2}~{\mathrm tr}~U_{C_{i}} )^j  ~~.
\end{equation}
The parameter of this action have 
been optimized in Ref. \cite{DeGrand96.2} and
are reproduced in Table \ref{tab:weights}.
\begin{table}[h]
\begin{center}
\begin{tabular}{|l|c|c|c|c|}
\hline
$w(i,j)$            & $j=1$   &  $j=2$  & $j=3$ & $j=4$   \\
\hline
\hline
$i=1$ (plaquettes)  & $.3333 $& $.00402$ & $.00674$ & $.0152$ \\
$i=2$ (6-link loops)& $.08333$& $.0156 $ & $.0149 $ & $-.0035$ \\
\hline
\end{tabular}
\end{center}
\caption{\sl Weight coefficients of the simplified fixed point action}
\label{tab:weights}
\end{table}

The notion
of blocking 
is recalled here 
in order to define some notations.
Fat links
$V_{x,\mu}$ are defined for $x_{\nu}= odd $ for all $\nu=1,2,3,4$ only.
These are the nodes of the blocked lattice.\footnote{The blocked lattice could 
be
defined as well shifted by one fine lattice step along one or more
directions.}
The blocking transformation expresses a fat link
in terms of the product of the two links along the fat link
plus a sum over 
six rectangular $4$-link staples
passing by the fat link:
\begin{eqnarray}\label{eq:blocking}
\tilde{V}_{x,\mu} & = & c_{1}^{block} ~ U_{x,\mu}U_{x+\hat{\mu},\mu}
      \nonumber \\
 & & + \sum_{\nu \neq \mu} c_{2}^{block} \left(
U_{x,\nu} U_{x+\hat{\nu},\mu} U_{x+\hat{\nu}+\hat{\mu},\mu}
U_{x+2\hat{\mu},\nu}^{+}  \right.  \\
 & &  ~ ~ ~ ~ ~ ~ ~ ~ +  \left. U_{x-\hat{\nu},\nu}^{+} U_{x-\hat{\nu},\mu}
 U_{x-\hat{\nu}+\hat{\mu},\mu}
U_{x-\hat{\nu}+2\hat{\mu},\nu} \right)   . \nonumber
\end{eqnarray}
The actual fat link $V_{x,\mu}$ is then obtained by normalization to
$SU(2)$:
\begin{equation}\label{eq:normalization}
V_{x,\mu} = {{\tilde{V}_{x,\mu}} \over
       {\sqrt{{\mathrm det}\left(\tilde{V}_{x,\mu}\right)}}}.
\end{equation}
The following blocking parameters have been recommended for this type of 
RG transformation
in Ref. \cite{HasenfratzNPB454I}
\begin{eqnarray}\label{eq:blockparam}
c_{2}^{block} & = & 0.12   \nonumber  \\
c_{1}^{block} & = & 1 - 6~ c_{2}^{block}   ~~.
\end{eqnarray}

\subsection{Inverse Blocking}

Inverse blocking consists in searching the constrained
minimum with respect to the fine
links $~U$ of an extended action
\begin{equation}\label{eq:inverse_blocking}
~~~  S_{ext}(U)=S_{FP}(U) + \kappa ~ T(U,V)  ~~,
\end{equation}
which includes, besides of the fixed point action $S_{FP}$, the blocking kernel
\begin{equation}\label{eq:TUV}
T(U,V) = \sum_{fat~ links}
\left(
\max_{W \epsilon SU(2)} \left( {\mathrm tr} \left( W^{+} \tilde{V} \right) \right)
                       - {\mathrm tr} \left( V^{+} \tilde{V} \right) \right)
                     \geq 0
\end{equation}
with $\kappa$ as a Lagrange multiplier. 
For better readability, we have dropped the labels $x,\mu$ of the 
fat links $V_{x,\mu}$ (describing the blocked configuration) and of the 
$\tilde{V}_{x,\mu}$. The latter are expressed in terms of the fine $~U$'s 
exactly as
in (\ref{eq:blocking}) and are not related to $V_{x,\mu}$ in the 
present context.

For comparison, cooling the gauge field configuration $~U$
means to minimize  
the new action with $\kappa=0$. 
The result of cooling is no more  
influenced by the coarse configuration $~V$ in the process of 
relaxation. The unconstrained
minimum
$~U$ is a classical solution on the lattice with
respect to $S_{FP}$. It can only 
depend on the initial configuration if that
determines a non--trivial basin of attraction.

Concerning the organization of measurements, we differ from DeGrand et al.
They have simulated on the coarse lattice, using the fine
lattice only for measuring the topological charge on the inversely blocked
configurations. The coarse lattice was their reference scale.
We carried out simulations on the fine lattice
(at some lattice scale $~a$) chosing a
$\beta$--value to produce a Monte Carlo ensemble of
gauge field configurations $U^{MC}$ according to the fixed point action. 
Most of these configurations have been used
only for comparison as described below. Another part of them 
(in a separate run with the same $\beta$) has been used
to produce an ensemble of
coarse lattice configurations $~V$ by blocking the fine lattice configurations.
Due to variance reduction it was possible to restrict this second ensemble 
to considerably less configurations than the other.
At scale $~2 \cdot a~$ (and bigger) the coarse ensemble is 
expected to describe the same physics as the fine ensemble.
We could have asked to which $\beta^{\prime}$ the coarse lattice
configurations correspond. 
However, for the sake of clarity in the comparison we always 
refer to the original $\beta$--value and $a$ as our reference scale.
The coarse lattice configurations $~V$ have been analyzed with respect to
their topological properties by inverse blocking, $~V \rightarrow U^{IB}$. 

To create
the coarse configurations by blocking from another equilibrium ensemble 
is no matter
of principle.
It is, however, an advantage to keep the Monte Carlo configuration on the fine
lattice in order to use it as a start configuration for the constrained
minimization.
Due to this particular circumstance the inversely blocked configuration 
$U^{IB}$ can be imagined as a smoothed copy 
of the original configurations $U^{MC}$.
Nevertheless, it can exhibit no other topological structure than present in
the blocked configuration $~V$ (which influences the relaxation
through the blocking functional $\kappa T(U,V)$). 
In what follows we shall denote the
combined application of one step of blocking followed by one step of
inverse blocking as {\it smoothing}.

Classical configurations are particular.  
Instanton
solutions on the fine lattice are always reproduced under smoothing. 
Their topological structure is the same on
the coarse and the fine lattices. 
Not only the action is recovered but also the shape of the
topological charge distribution and the location of monopole currents.
We have produced
a set of almost classical configurations
by cooling from hot starts
using the simplified fixed point action (\ref{eq:fp_action}).
This set has then been used for testing our code of
the smoothing
algorithm.
If one blocks a classical solution, one finds $S_{FP}(U)=S_{FP}(V)$.

However, this technique is intended to be 
applied to non--classical configurations.
Before entering our investigations, we
have tested, creating fine lattice configurations $U^{MC}$ 
(Monte Carlo generated with $S_{FP}$ at various $\beta$--values) 
and blocking them into
coarse configurations $~V$,
{\it which} $\kappa$--value must be used in the extended action
(\ref{eq:inverse_blocking}) 
driving $~U$ from $U^{MC}$ to $U^{IB}$.
The relaxed configurations $U^{IB}$ should turn 
the inequality (\ref{eq:inequality1}) into an equality 
relating $U^{IB}$ to $~V$.
Under these circumstances we were not able
to confirm the 
Lagrange multiplier $\kappa=12$ 
recommended in Ref. \cite{HasenfratzNPB454I} 
(for the given type of RG transformation in $SU(3)$).
Instead, we have found that
a choice $\kappa=5.15$
(slightly depending on $\beta$)
is suitable,
throughout the $\beta$ range under study, to guarantee the
equality in (\ref{eq:inequality1}) within
an accuracy of one to three per cent.
We have adopted 
the fixed point action but decided to  
use our empirical value for $\kappa$. 
In principle, 
$\kappa$ and the parameters of the
fixed point action are closely tied together.
They
should be optimized w. r. t. (\ref{eq:inequality1}) 
using an ensemble
of configurations created with exactly this action. This is a much more
ambitious project than it has been pursued until now. The authors of  
Ref. \cite{DeGrand96.1} are continuing this 
program.\footnote{private communication}
This whole discussion puts a question mark on the exact
relation between the action and $\kappa$.
Also in view of this, we consider the results of the 
present paper as 
exploratory. 

Knowing the extended action (\ref{eq:inverse_blocking}) we can distinguish
three types of links on the fine lattice.
Links along the fat links 
and links lying inside coarse pla\-quettes
contribute to $\tilde{V}$ and are 
updated (relaxed) with respect to $S_{ext}$.
All other fine links are not subject to constraints and
are updated
with respect to $S_{FP}$ alone.
\begin{figure}[!thb]
\centering
\epsfig{file=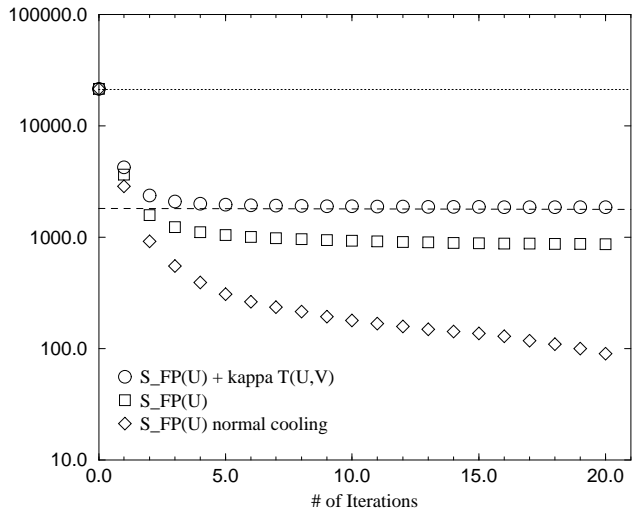,width=9.5cm,angle=0}
\caption{\sl Relaxation towards the inversely blocked configuration
vs. cooling for a typical Monte Carlo configuration generated at
$\beta=1.5$ on a $12^3\times 4$ lattice}
\label{fig:relax}
\end{figure}
A typical relaxation process is shown in Fig. \ref{fig:relax}. The
upper curve demonstrates how the
extended action $S_{FP}(U)+ \kappa ~ T(U,V)$ decays versus number of iterations.
The upper horizontal line marks the action of the original Monte Carlo
generated configuration $S_{FP}(U^{MC})$, the lower horizontal
the action of the blocked
configuration $S_{FP}(V)$. After $~V$ has been obtained  
from $U^{MC}$ by blocking and one has plugged in $U^{MC}$ into the
extended action $S_{ext}(U)$,
$T(U,V)$ vanishes by construction. Therefore the minimization of $S_{ext}$
starts at the level of $S_{FP}(U^{MC})$.
After the constrained minimum configuration 
$U^{IB}$ has been found, 
the extended action is shared between the fixed point
action $S_{FP}(U^{IB})$ and the functional $\kappa ~ T(U^{IB},V)$.
In Fig. \ref{fig:relax} the history of $S_{FP}(U)$ is separately recorded 
by the curve in the middle. The lowest curve shows 
how
$S_{FP}(U)$ would evolve
under cooling
(with
$\kappa$ put to zero). 
After $20$ iterations the unconstrained 
relaxation runs into configurations $~U$
with
a fixed point action $S_{FP}(U)$ one order of magnitude below
$S_{FP}(U^{IB})$ and the action decays further. On the other hand, this
demonstrates that inverse blocking does not need huge numbers of 
iterations to achieve convergence if
a good start configuration is known. 
A particular relaxation scheduling, 
for instance a modification of the
action during the first iterations, is not necessary.

\section{String Tension at $T=0$ and Finite \\
 Temperature}

The inverse coupling
constant $\beta$ enters through the Gibbs weight
\begin{equation}\label{partition_function}
Z = \int DU \exp(- \beta S_{FP}(U))
\end{equation}
into the creation of fine lattice configurations.
It defines the scale through the lattice spacing $a(\beta)$.
We wanted to compare the naive charge with 
L\"uscher's charge which is, similar to the fixed point action,
very CPU intensive. Therefore we
have restricted ourselves
to a fine lattice of size $12^3\times 4$
for the finite temperature part of 
these explorative investigations. 

For $L_{t}=4$ the deconfinement transition was known to occur at
$\beta_c = 1.575(10)$ \cite{DeGrand96.2}.
This has been established by the crossing of the Binder cumulants
of the Polyakov line for several spatial lattice sizes.
We have not attempted here 
to improve the localization of the deconfining transition
or to do a finite size analysis.
Rather than doing that, we have simulated at $\beta$--values safely lower
($\beta=1.40$, $\beta=1.50$ and $\beta=1.54$) and higher
($\beta=1.61$ and $\beta=1.80$) than the
deconfinement transition.

\subsection{Lattice Spacing from the Zero Temperature \\
String Tension}

In order to measure the lattice spacing $a$ at each of the
$\beta$--values
to be used in the finite temperature simulations
we have complementarily measured 
the zero temperature string tension
on a $12^4$ lattice.
No topological features have 
been analysed in this part of our calculations. 
\begin{table}[h]
\begin{center}
\begin{tabular}{|l|c|c|c|c|c|}
\hline
$\beta$     & $1.40$    & $1.50$     & $1.54$     & $1.60$     & $1.80$      \\
\hline
\hline
Monte Carlo  & $0.445(8)$ & $0.245(3)$ & $0.193(3)$ & $0.124(2)$ & $0.0380(9)$ \\
inv. blocked & $0.228(4)$ & $0.182(1)$ & $0.156(3)$ & $0.108(3)$ & $0.0299(1)$ \\
\hline
\hline
Monte Carlo  & $152(3)$   & $135(2)$   & $129(2)$   & $111(2)$   & $91.7(19)$  \\
inv. blocked & $77.9(10)$ & $100.5(10)$ & $104.5(20)$ & $96.8(20)$  & $71.2(20)$   \\
\hline
\end{tabular}
\end{center}
\caption{\sl String tension $\sigma ~a^2$ (above) and
                      $\sigma/\Lambda_{L}^2$ (below)
          at zero temperature
          from fuzzy Wilson loops in axial gauge
          on a $12^4$ lattice
            (with two--loop expression for $a(\beta)\Lambda_{L}$)
}
\label{tab:loop_string_tens}
\end{table}
The string tension has been obtained for Monte Carlo 
configurations in the axial gauge
from smeared Wilson loops. The same has been measured on smoothed
configurations for comparison.
The axial gauge is enforced in chronological order
time slice by time slice
(except for the last one)
after a certain direction has been chosen
as Euclidean time. This is done 
putting each temporal link $U_{4}(x)$ 
equal to unity by applying a suitable gauge transformation
$g(x+\hat4)$ at the end of that link.
Smearing \cite{Smearing} has been applied to configurations in this gauge.
This
is an iterative operation
(without change of scale)
replacing each spatial link $~U$ by
$U^{smear}$ through a procedure similar
to (\ref{eq:blocking}) 
(since without change of scale, 
with only one step in $\mu=1,2,3$ direction
and with only spatial staples
involved, $\nu=1,2,3~~\nu \neq \mu$).
In the axial gauge, timelike parts of the Wilson loops are mostly equal 
to unity 
(if they do not contain 
links from the last to the first time slice). Smearing effectively replaces
the fixed--time parts of planar Wilson loops by 
weighted sums over paths, running from the 
quark to the antiquark site (a distance $R$ apart).
The temporal extension
of the Wilson loop is denoted as $T$.
In view of the maximal spatial distance $R=6$
of quark and antiquark,
we have applied $N_{smear}=6$ smearing iterations with
an optimal smearing parameter (the notation is as in
(\ref{eq:blocking}) )
\begin{eqnarray}\label{eq:smearparam}
c_{2}^{smear} & = & 0.125 \nonumber \\
c_{1}^{smear} & = & 1 - 4~ c_{2}^{smear}  .
\end{eqnarray}
This parameter had been optimized in order to get an early plateau
in $T$
of $\log(W(R,T+1)/W(R,T)$
for the Monte Carlo configurations
at $\beta=1.5$. For simplicity, it
has been applied in all measurements of the zero 
temperature string tension.

The potential is fitted (for all $R$) from exponential fits
over the range $T=2...5$. For the inversely blocked configurations we have
restricted
the fit range to $T=3...5$. The potential $V(R)$ has been fitted with
a $3$-parameter fit that includes a constant, Coulomb and linear part over
the range $R=1...5$.
For the inversely blocked 
configurations the Coulomb part is consistent with zero.
The string tension is obtained from the linear part of the potential.
The $\beta$ dependence and the comparison 
between normal Monte Carlo configurations
and inversely blocked (smoothed) configurations 
is presented in Table \ref{tab:loop_string_tens}.

The string tension for inversely blocked configurations
has a scaling window
(assuming asymptotic scaling
with the two--loop $a(\beta)$) including our measurements from $\beta=1.5$ to
$\beta=1.6$. 
The string tension measured at $\beta=1.8$
suffers from a small volume effect, since the spatial box size
at this $\beta$ is physically not larger than
$2.35/T_c$.
For definiteness, we choose the string tension from the Monte Carlo 
configurations
in order to express non--perturbatively
$a(\beta)$ and the
$4$-volume of the lattice
through the invariant
string tension.

According to the RG philosophy 
the same large scale physics should be represented by 
the coarse configurations (if obtained by blocking) and 
the original Monte Carlo configurations (on the fine lattice), 
as long as the blocking scale is still safely separated from the
physical scale 
under investigation. It is not clear a priori
that the inversely blocked
configurations $U^{IB}$ still contain  
the same large scale physics. In case of the string tension
this could only be the case if
it decouples from the
ultraviolet fluctuations.
In order to test this 
it makes sense to compare the string tensions of Monte Carlo
and inversely blocked configurations at various $\beta$.
The string tension is built up at large distances 
and is expected to coincide for both types of configurations, 
while the
perturbative Coulomb part of the potential is not expected 
to survive inverse blocking. We compare in table
\ref{tab:loop_string_tens} the zero temperature string tension as found on 
Monte Carlo and inversely blocked configurations, both on the fine lattice.
We find our expectation 
largely confirmed.
The string tension of inversely blocked configurations is somewhat
smaller 
($75$ to $90$ per cent depending on $\beta$) than the string tension of
normal Monte Carlo configurations. 

The exception at the lowest $\beta=1.4$ can be explained by the fact
that at very strong coupling the blocking scale interferes with the
confinement scale. This is supported by the observation that also
the topological structure is not truly captured. This becomes obvious
if one shifts 
the coarse lattice (in
one of the $2^4-1$ directions) relative to the fine lattice.
This is the reason why we must exclude the lowest $\beta$-value also from
extracting the temperature dependence of the topological
susceptibility.

This is place to compare cooling and inverse
blocking as far as the string tension is concerned.
Cooling 
conserves the string tension \cite{DiGiacomoString} as a function of
the cooling steps only 
as long as the topological
susceptibility goes through a plateau.
The topological susceptibility is
{\it defined} as the plateau value in this method \cite{DiGiacomoTopcool}. 
Cooling as used by the MIT group \cite{Negele94} in order to
measure hadronic correlation functions goes much farther 
and does not conserve the string tension. 

\subsection{Is the $T\ne0$ String Tension Changing under \\ 
Inverse Blocking ?}

We discuss now our results concerning the temperature
dependent string tension $\sigma(T)$. Although this is not our
main interest, we
want to see to what extent this quantity
is also insensitive to inverse blocking. 
The configurations are Monte Carlo generated 
on asymmetric
lattices of size $12^3\times 4$. The
string tension $\sigma(T)$ should vanish in the deconfinement
phase, at $\beta > \beta_c$.
A suitable definition of the string tension for
thermal lattices is provided by the correlator of
Polyakov lines
\begin{equation}\label{eq:Polyakov}
L({\bf x}) = {\mathrm tr} \prod_{t=1}^{L_t} U_{\bf{x}t,4}
\end{equation}
which, in the confinement phase, exhibits the confining
potential in the form
\begin{equation}\label{eq:stringtension}
\langle L({\bf 0}) L({\bf x}) \rangle
\propto \exp (- V_{\overline{q} q}^{conf}({\bf x})/T)
\end{equation}
where $V_{\overline{q} q}^{conf}({\bf x}) \sim \sigma(T) |{\bf x}|$ at
large distances. $V_{\overline{q} q}^{conf}({\bf x})$
contains
the self--energies of $q$ and $\overline{q}$ and the perturbative
Coulomb potential, too. These are expected to change under
inverse blocking.

\begin{figure}[!thb]
\centering
\epsfig{file=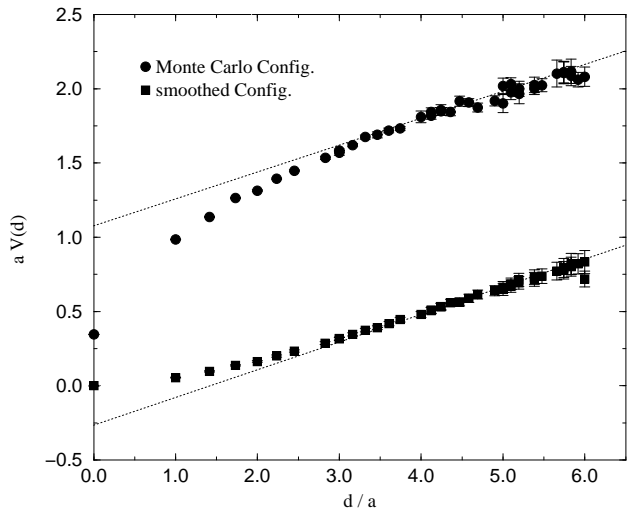,width=9.5cm,angle=0}
\caption{$\overline{q} q $ \sl potential from Polyakov line correlators
for Monte Carlo (above) and inversely blocked configurations
at $\beta=1.5$ in the confinement phase}
\label{fig:potential}
\end{figure}
In the measurement of Polyakov lines and their correlators
there is no place for smearing and no sense in performing the axial gauge.
The normalized correlators in the Monte Carlo
and the inversely blocked ensemble
can be
\begin{table}[h]
\begin{center}
\begin{tabular}{|l|c|c|c|}
\hline
$\beta$     & $1.40$      &  $1.50$    & $1.54$ \\
\hline
$T/T_{c}$ & $0.656$     &  $0.834$   & $0.919$ \\
\hline
\hline
Monte Carlo  & $0.253(23)$ & $0.193(4)$ & $0.176(5) $ \\
inv. blocked & $0.242(8) $ & $0.189(2)$ & $0.137(3) $ \\
\hline
\hline
Monte Carlo  & $86(8)$     & $110(4)$   & $118(4)$ \\
inv. blocked & $83(3)$     & $104(2)$   & $92(2)$ \\
\hline
\end{tabular}
\end{center}
\caption{\sl String tension $\sigma(T) ~a^2$ (above) and
                      $\sigma(T)/\Lambda_L^2$ (below)
                      from Polyakov line
correlators at various temperatures $T < T_{c}$
in the confinement phase
(with two--loop expression for $a(\beta)\Lambda_{L}$)
}
\label{tab:pol_string_tens}
\end{table}
directly compared for the three $\beta$--values that belong to 
the confinement phase.
Fig. \ref{fig:potential} shows for $\beta=1.5$ the logarithm of the
(suitably normalized) Polyakov line correlator as function of the distance.
In Table \ref{tab:pol_string_tens} we collect
the extracted values of the
string tension at the three temperatures in the confinement phase
for Monte Carlo and inversely blocked configurations.
There is an interesting temperature dependence of the thermal string
tension, but no big difference between the string tensions
measured on the Monte Carlo and the inversely blocked ensemble. 
The reduction is less than $5$ per cent at temperatures
$T < 0.8~T_{c}$ and grows to 
more than $20$ per cent at $T > 0.9~T_{c}$.
We observe also at $T<T_c$ that the 
original Monte Carlo string tension is practically the same
as the string tension measured on the much
smoother inversely blocked configurations and seems to be
independent of
the short range fluctuations.

\section{Gross Topological Features Near to the \\
 Deconfinement Transition}

The first measurement of topological features
of pure $SU(2)$ gauge theory with the new method
concerns
the temperature dependence of
the topological
susceptibility of pure $SU(2)$ gauge theory, here over the 
range
from $0.8~T_c$ to $1.7~T_c$.
This question
has been studied already in Ref. \cite{DeGrand96.2} on lattices
of invariable spatial size (both in lattice units $L_s$ and physical units) by
varying the temporal extent $L_t$.
We keep the lattice size invariantly equal to 
$12^3\times 4$ while varying $\beta$.
This additional measurement might be of interest as such since
now the spatial size changes with the temperature.

\subsection{Topological Susceptibility Across the Deconfining \\
Phase Transition}

We have measured on Monte Carlo configurations (separated by $20$ updates)
on the lattice $12^3\times 4$ the topological charges
according to L\"uscher's 
definition, $Q_{Luescher}$\cite{Luescher,LuescherImplem},
and using the naive topological
density \cite{DiVecchia}. 
The latter is defined (actually in a symmetrized way) in terms of plaquettes
\begin{equation}\label{eq:q_naive}
q_{naive}(x) = - \frac1{2^9 \pi^2} \sum_{\mu,\nu,\sigma,\rho=-4}^{+4}
\epsilon_{\mu\nu\sigma\rho}
{\mathrm tr} \left(U_{x,\mu,\nu} U_{x,\sigma,\rho}\right)
\end{equation}
with
\begin{equation}\label{eq:Q_naive}
Q_{naive} = \sum_{x} q_{naive}(x)
\end{equation}
(summed over lattice points). The statistics of analyzed Monte Carlo
configurations amounts to $300$ configurations.
These measurements will be confronted for each $\beta$ with
corresponding ones on an independent ensemble of $100$
inversely blocked (smoothed) configurations.
There is no room for details of
L\"uscher's charge \cite{LuescherImplem}.
Important for us is 
that it is also written as a sum of localized contributions
\begin{equation}\label{eq:Q_luescher}
Q_{luescher} = \sum_{x} q_{luescher}(x)
\end{equation}
(summed over hypercubes).
The time consuming part
of this algorithm is the numerical integration of $2$- and 
$3$-dimensional integrals. It should be encouraging that
evaluating
the topological charge of 
inversely blocked configuration needs only $\frac{1}{10}$ of CPU time 
(function calls) compared to Monte Carlo configurations. 
The topological susceptibility is defined as
\begin{equation}
\chi_{top}= {{\langle Q^2 \rangle} \over {N_{sites} a^4}}  ,
\end{equation}
with total charges $Q$
corresponding here always
to the naive and L\"uscher's charge.
$N_{sites}$ is the number of
lattice points $L_{s}^3 \times L_{t}$ and we estimate the physical
lattice volume $N_{sites} a^4$ with the
help of the zero temperature string tension.
Data for Monte Carlo and inversely blocked 
configurations as function of $T/T_{c}$
(yet without renormalization of the naive topological susceptibility)
are collected in Table \ref{tab:suscept}.
\begin{table}[h]
\begin{center}
\begin{tabular}{|l|c|c|c|c|}
\hline
$\chi_{top}$& $\beta$ & $T/T_{c}$ & $\chi_{top}^{luescher}/\Lambda_L^4$ & $\chi_{top}^{naive}/\Lambda_L^4$ \\
\hline
\hline
Monte Carlo  & $1.40 $ & $0.656$ & $0.224(17) 10^4$ & $0.146(12) 10^2$ \\
             & $1.50 $ & $0.834$ & $0.287(24) 10^4$ & $0.370(27) 10^2$ \\
             & $1.54 $ & $0.919$ & $0.310(27) 10^4$ & $0.449(43) 10^2$ \\
             & $1.61 $ & $1.089$ & $0.235(19) 10^4$ & $0.753(59) 10^2$ \\
             & $1.80 $ & $1.732$ & $0.190(16) 10^4$ & $0.308(24) 10^3$ \\
\hline
inv. blocked & $1.40 $ & $0.656$ & $0.127(17) 10^3$ & $0.84(11) 10^2$  \\
             & $1.50 $ & $0.834$ & $0.358(47) 10^3$ & $0.240(29) 10^3$ \\
             & $1.54 $ & $0.919$ & $0.324(43) 10^3$ & $0.239(30) 10^3$ \\
             & $1.61 $ & $1.089$ & $0.138(25) 10^3$ & $0.962(18) 10^2$ \\
             & $1.80 $ & $1.732$ & $0.164(11) 10^2$ & $0.100(70) 10^2$ \\
\hline
\end{tabular}
\end{center}
\caption{\sl Topological susceptibilities in the Monte Carlo and inversely 
blocked samples (with two--loop expression for $a(\beta)\Lambda_{L}$)
}
\label{tab:suscept}
\end{table}
Only the topological susceptibility $\chi_{top}^{luescher}$ 
of the inversely blocked ensemble can be interpreted 
as a topological susceptibility
(in the sense of the classically perfect topological charge).
We have to agree that the topological charges belong to the
coarse lattice and are defined at 
a scale given by the coarse lattice spacing.
The apparent topological susceptibility evaluated on the
Monte Carlo ensemble of fine lattice configurations 
is bigger than $\chi_{top}^{luescher}$ 
by one order of magnitude (in the confinement
phase) und up to two orders of magnitude (in the deconfinement phase).
This does not mean that the enormous difference must be attributed 
to {\it physical} fluctuations on the scale of the fine lattice spacing.
Their contribution can 
be quantified only if one inversely blocks down to the
next finer lattice (of size
$24^3 \times 8$) with lattice spacing $\frac12 \cdot a$. 
Even lacking this information, the comparison shows
how much the measurement of the topological susceptibility
without the step of inverse blocking fails.
This means in particular 
that the (simplified) fixed point action alone does not prevent 
dislocations from contributing to $\chi_{top}$.

Let us now discuss to what extent the naive topological charge density operator
can replace L\"uscher's construction in order to extract the continuum
topological charge density of a lattice configuration. 
It is known for simulations with Wilson's action
in the range $\beta=2.$ to $3.$, that the corresponding
perturbative renormalization factor \cite{Renormalization}
is very small compared to one. 
Also for simulations with the fixed point action, the 
topological susceptibility, 
evaluated immediately for Monte Carlo configurations
in the confinement phase, 
is two
orders of magnitude smaller with the naive charge than
the susceptibility defined
by means of  
L\"uscher's charge $Q_{luescher}$ .

\begin{figure}[!thb]
\centering
\epsfig{file=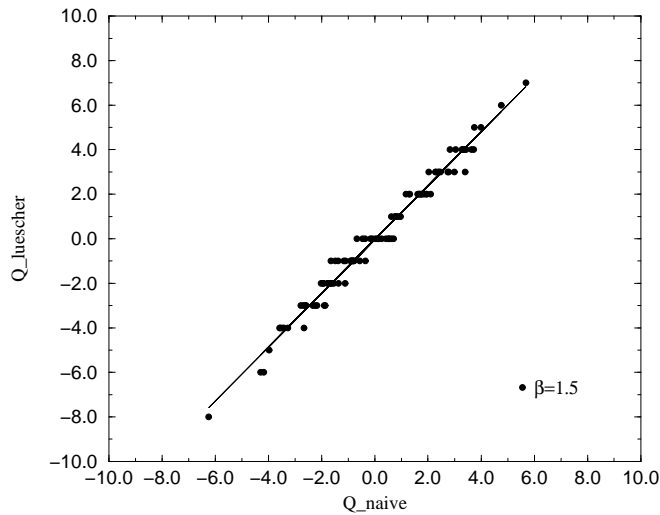,width=9.5cm,angle=0}
\caption{\sl Scatter plot of L\"uscher's vs. naive
topological charge for inversely blocked configurations at $\beta=1.5$}
\label{fig:Z_Faktor_scatter}
\end{figure}

In contrast to this, 
the susceptibilities $\chi_{top}^{naive}$ and $\chi_{top}^{luescher}$
differ by less than $30$ per cent
for the inversely blocked ensemble 
of configurations. 
The remaining discrepancy
can be illustrated with
Fig. \ref{fig:Z_Faktor_scatter}. This 
scatter plot of inversely blocked configurations shows how
$Q_{luescher}$ is correlated
with $Q_{naive}$.
There is only a small variance between the two charges 
evaluated after one step of inverse
blocking. The naive charge differs from L\"uscher's charge never
by more 
than one unit. We can 
define a renormalization factor for the naive charge
$Z_q^{(1)}(\beta)$ on inversely blocked configurations 
by the slope of the scatter plot.
The corresponding
numbers are collected in Table \ref{tab:Zfactor} for our set of $\beta$--values
near to the phase transition.\footnote{For our highest 
$\beta$ non--zero
charges are very rare, such that the $Z_q^{(1)}$
factor is not reliably known.} 
The renormalization factor 
can be used to relate the continuum to the naive topological
density after the first step of inverse blocking:
\begin{equation}\label{eq:Zfactor}
Z_q^{(1)} q_{cont}(x)~a^4 = q_{naive}^{(1)}(x)~a^4 ~~.
\end{equation}
\begin{table}[h]
\begin{center}
\begin{tabular}{|l|c|c|c|c|}
\hline
$\beta$             & $1.50$ & $1.54$ & $1.61$ & $1.8$  \\
\hline
\hline
$Z_q^{(1)}(\beta)$    & $.826$ & $.877$ & $.843$ & --  \\
\hline
\end{tabular}
\end{center}
\caption{\sl Effective renormalization $Z_q^{(1)}$ factors of the
naive topological charge
density for one step of inverse blocking 
at various values of $\beta$ on the $12^3\times 4$ lattice}
\label{tab:Zfactor}
\end{table}

Further steps
of inverse blocking will bring $Z_q^{(n)}(\beta)$ nearer to one.
Compared to the renormalization factor for Monte Carlo configurations,
this effect is similar to the effect of improving the operator
of topological
density \cite{ImprovedQ} proposed by Di Giacomo et al. 
These operators are constructed 
exactly as in (\ref{eq:q_naive}) 
in terms of suitably smeared links 
$U^{(n)}$ 
(iteratively smeared to the $n$-th level). This or other improved
operators of topological density \cite{VanBaal,DeForcrand96,DeForcrand97} 
can be used for
analysing inversely blocked configurations.
For the topological susceptibility the renormalization
factor as derived from the scatter plot is sufficient.
\begin{figure}[!thb]
\centering
\epsfig{file=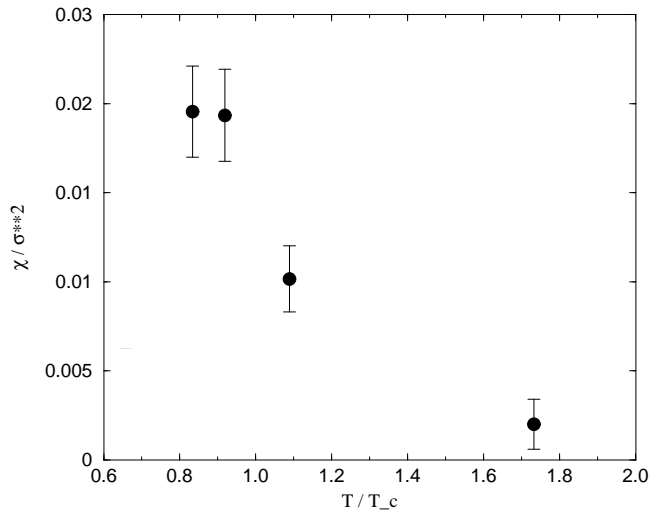,width=9.5cm,angle=0}
\caption{\sl Topological susceptibility for inversely blocked configurations
across the deconfining phase transition for
L\"uscher's
topological density}
\label{fig:suscept_ib}
\end{figure}
When the naive charges (that went into the susceptibility of smoothed
configurations in Table \ref{tab:suscept}) 
are multiplicatively corrected by 
the $Z_q^{(1)}(\beta)$ factors from Table \ref{tab:Zfactor}
we get a unique $\beta$-dependence of $\chi_{top}$. We have
shown in Fig. \ref{fig:suscept_ib} the 
topological suceptibility for our four temperatures.
The lattice
spacing $a(\beta)$ has been non--perturbatively
expressed through the zero temperature
string tension as measured
at each respective $\beta$--value 
on the $12^4$ lattice.

Assuming the value 
$\sigma=(440 {\mathrm MeV})^2$ 
for the zero temperature string tension  
we estimate 
at $T=0.834~T_c$ a susceptibility $\chi_{top}=(165.5 {\mathrm MeV})^4$.
Compared with the standard value of $\chi_{top}=(180 {\mathrm MeV})^4$
at zero temperature
there is not much room for increase at smaller temperature.
\begin{figure}[!thb]
\centering
\epsfig{file=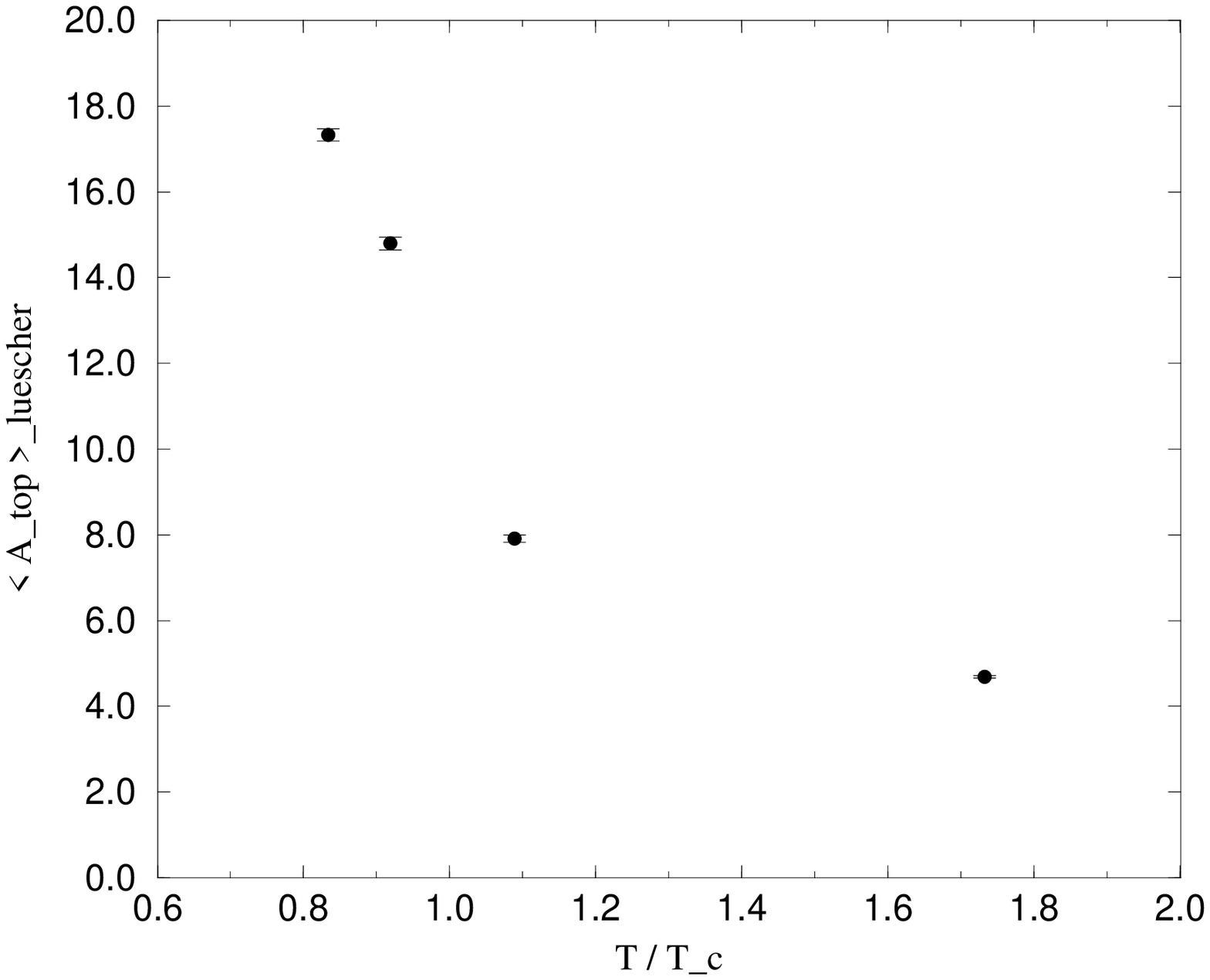,width=9.5cm,angle=0}
\caption{\sl Average topological activity $\langle A_{top} \rangle$
(according to L\"uscher's charge)
of inversely blocked configurations
as function of $T/T_{c}$}
\label{fig:A_t}
\end{figure}

For building instanton models of vacuum structure 
it is important to know not only the
topological susceptibility but also the average density of instantons and 
antiinstantons. Only in the most simple-minded dilute gas model
the topological susceptibility coincides with this density.
In addition to the
total charge (eqs. (\ref{eq:Q_naive}) or (\ref{eq:Q_luescher})),
the following operator is tentatively chosen
to represent 
$N_{+} + N_{-}$,
the number of instantons plus antiinstantons
\begin{equation}\label{eq:A_t}
A_{top} = \sum_{x} |q(x)|  ~~~.
\end{equation}
We can understand this,
somewhat loosely,
as a measure for the temperature dependent average glueball field
(gluon condensate).
Let us call it
''topological activity''. Compared with measurements of this quantity on
Monte Carlo configurations, it is reduced 
by $\frac{1}{25}$ to $\frac{1}{40}$ on inversely blocked configurations,
depending on temperature. For normal Monte Carlo configurations
there is too much noise
at low values of $|q(x)|$.  Without inverse blocking,
$A_{top}$ is unreliable as an
estimator of the number of instantons plus antiinstantons. 
In Fig. \ref{fig:A_t}
we show the average topological activity of inversely blocked configurations as
function of $T/T_{c}$. Between $\beta=1.54$ and $\beta=1.61$ (over temperatures from
$T=0.92~T_{c}$
to $T=1.09~T_{c}$)
it drops by $50$ per cent, but the decrease becomes slower
at higher temperature.
\footnote{In contrast to that, the topological activity measured on
Monte Carlo configurations is less 
suppressed as a function of temperature.}

\begin{figure}[!thb]
\centering
\epsfig{file=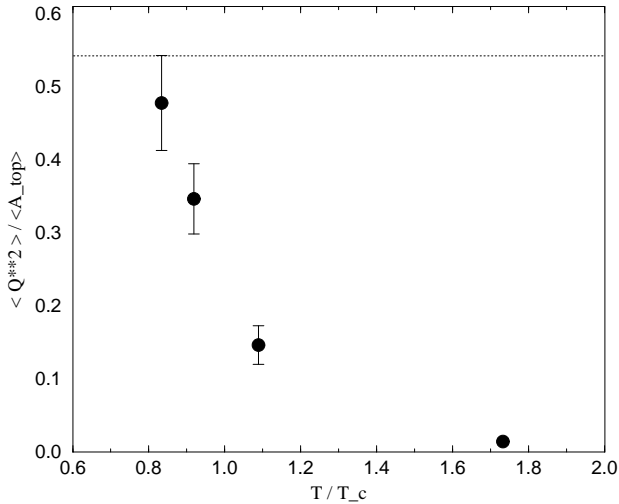,width=9.5cm,angle=0}
\caption{\sl Ratio of topological susceptibility $\chi_{top}$ to
average topological activity $\langle A_{top} \rangle/N_{sites} a^4$
(according to L\"uscher's charge)
of inversely configurations
as function of $T/T_{c}$}
\label{fig:QsqdurchA}
\end{figure}
Concentrating now on the inversely blocked configurations,
the topological susceptibility is strongly suppressed 
with increasing temperature,
compared with the decrease of the
topological activity. 
Fig. \ref{fig:QsqdurchA} demonstrates this 
additional suppression of uncompensated 
topological charge in the deconfinement
phase. Apart from this observation,
an interesting low temperature bound is suggested (see the dotted horizontal
line in Fig. \ref{fig:QsqdurchA}) :
\begin{equation}\label{eq:1/N}
\chi_{top} \leq \frac{6}{11} \frac{\langle A_{top} \rangle}
                           {N_{sites} a^4}
               =  \frac{6}{11} \langle n_{+} + n_{-} \rangle  .
\end{equation}
There is a factor $\frac{4}{b}=\frac{12}{11N_c}$ (equal to the entropic bound)
in this relation where
$b$ is the one--loop coefficient in the QCD $\beta$-function.
Models describing this $O(1/N_c)$ suppression of the topological
susceptibility compared to the densities $n_{\pm}$
of instantons and antiinstantons
have been discussed 
in an instanton liquid model
with scale invariant hard core interaction \cite{1/N} and in
a similar droplet  
model \cite{Muenster}.

We have found that  
({\it i}) the topological
susceptibility for smoothed configurations is of reasonable size
on the confinement side of the transition,
({\it ii}) the topological
susceptibility is smaller by a factor $\frac{4}{b}$ than
the topological activity at low temperature 
and that ({\it iii}) the
topological susceptibility decreases more than the topological activity
entering the deconfinement phase.

\subsection{Monopole Content of Monte Carlo and Inversely \\
Blocked 
Configurations}

The Abelian monopole degrees of
freedom
are exposed by putting the lattice configurations into the
maximally Abelian gauge \cite{Kronfeld}. Only few investigations
have addressed the question how the dynamics of Abelian monopoles
gets modified at the deconfinement
transition and within 
the deconfined phase \cite{Bornyakov2}. Abelian dominance
is expected to
become manifest at large distances. Hence it is natural to
study these aspects in the ensemble of inversely blocked configurations
which are smoothed at small distances.

It is not necessary to recall here details on
the Abelian projection \cite{Kronfeld}. By a gauge cooling algorithm
configurations are iteratively put into this gauge, 
whether they are directly Monte Carlo generated or obtained
by inverse blocking. Finally, each
configuration is mapped to a corresponding Abelian
one. The links are represented by $U(1)$ phase factors and the
non--diagonal fields are dropped.
Using this new ensemble of configurations, confinement can be studied
either in terms of Abelian Wilson loops formed out of
the Abelianized links or, alternatively, in terms of the
monopole currents. These are detected \cite{DeGrandToussaint} seeking for 
magnetic flux penetrating
the surface of a $3$-dimensional cube. Each time a non--zero flux is
detected,
the dual (with respect to the cube) link $x,\mu$ is said to carry
a monopole current $m_{\mu}(x)$.
\begin{figure}[!thb]
\centering
\epsfig{file=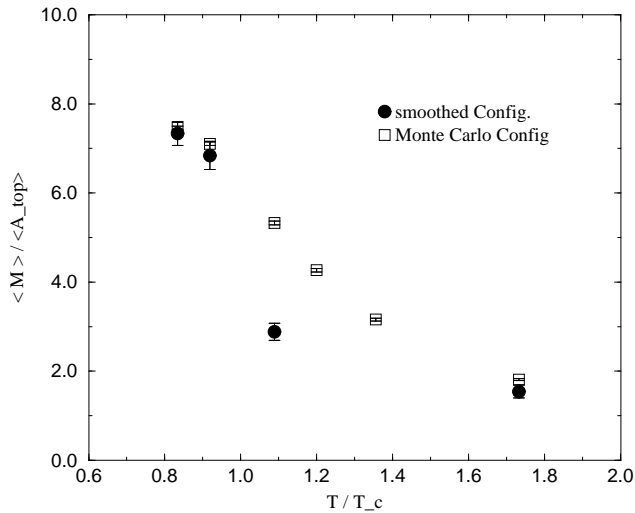,width=9.5cm,angle=0}
\caption{\sl Ratio between the averages of total monopole current number $M$
and topological activity $A_{top}$ as function of $T/T_{c}$ for Monte Carlo
and inversely blocked configurations}
\label{fig:MdurchA}
\end{figure}
We do not discuss here the correct description
of monopole condensation \cite{MonopolCondensation}
and restrict ourselves to
measurements
of the monopole
current numbers
\begin{equation}\label{eq:M_total}
M_{\mu}= \sum_{x} |m_{\mu}(x)| ~~.
\end{equation}
It is important to distinguish
spacelike ($\mu=1,2,3$) and timelike ($\mu=4$) dual links carrying monopole
currents. 
We will see that their numbers behave 
differently at the deconfinement temperature.
\begin{figure}[!thb]
\centering
\epsfig{file= 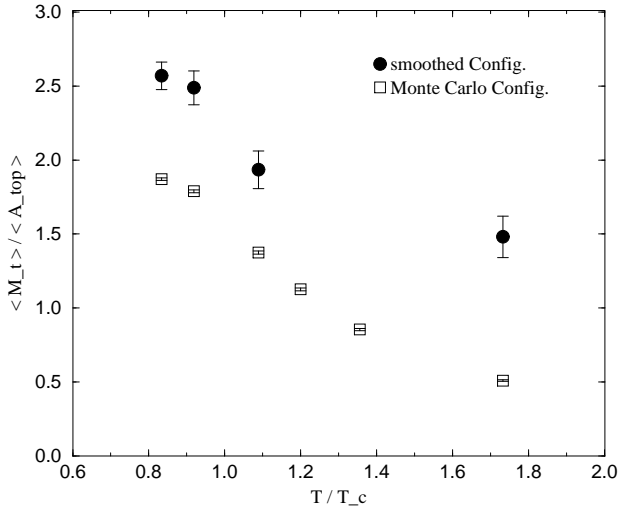,width=9.5cm,angle=0}
\caption{\sl Ratio between the averages of timelike monopole current number
$M_{t}$
and topological activity $A_{top}$ as function of
$T/T_{c}$ for Monte Carlo
and inversely blocked configurations}
\label{fig:MTdurchA}
\end{figure}

There are indications that the monopole current number
$M = \sum_{\mu} M_{\mu}$
is related to the topological activity $A_{top}$. Both quantities are
reduced by the same factor $\frac{1}{25}$ to $\frac{1}{40}$,
at the lowest and the highest temperatures, respectively,
as the result of blocking followed by inverse blocking. 
We show in Fig. \ref{fig:MdurchA} the ratio
$\langle M \rangle/\langle A_{top} \rangle$.
This ratio
is a smoothly falling function of temperature across the phase transition
for normal Monte Carlo configurations.
For inversely blocked configurations, however, we find an sudden 
drop of this ratio at $T_c$. This is mainly due to the suppression of
spatially directed monopole currents. The corresponding ratio for
timelike monopole currents alone (see Fig. \ref{fig:MTdurchA})
is a smooth function of temperature for Monte Carlo {\it and}
inversely blocked configurations.

\begin{figure}[!thb]
\centering
\epsfig{file=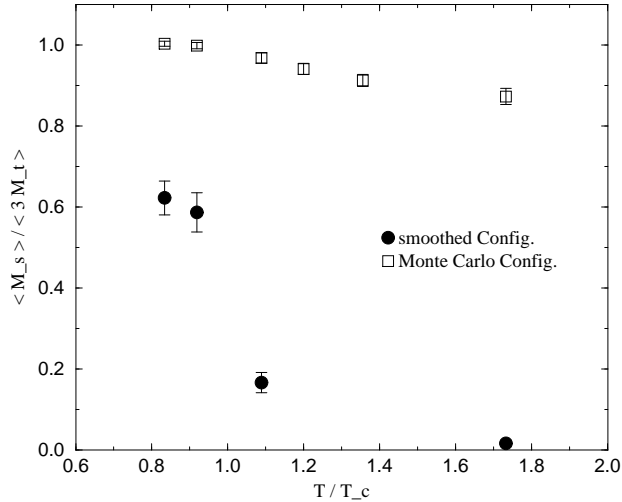,width=9.5cm,angle=0}
\caption{\sl Ratio between the averages of timelike and spacelike
monopole currents across the
deconfining phase transition for Monte Carlo and inversely blocked configurations}
\label{fig:monasymmetrie}
\end{figure}

The ratio between $M_s$ (summed over $\mu=1,2,3$)
and $3~M_t$ ($\mu=4$) is expected to change at the deconfinement
temperature \cite{Bornyakov2}. 
In Fig. \ref{fig:monasymmetrie}
the ratio
$\langle M_s \rangle/3 \langle M_t \rangle$
is shown for Monte Carlo and
inversely blocked configurations as function of $T/T_c$.
For the original Monte Carlo
configurations the monopole current number is space--time symmetric
in the confinement phase and the ratio
$\langle M_s \rangle/3 \langle M_t \rangle$
changes only slightly
over the temperature range interpolating from the confinement to
the deconfining phase. 
The deconfinement effect is not strong because it
is hidden by the
huge overall monopole activity in the Monte Carlo
configurations. The effect becomes clear only for inversely 
blocked configurations.
There exists an surplus of timelike
monopole currents already in the confinement phase (which probably must be
interpreted as a finite $L_{t}$ effect produced in the
process of blocking). There is a strong reduction of spacelike monopoles 
at $T=1.089 T_c$, and deeper in the deconfined phase 
they are completely suppressed.
\begin{figure}[!thb]
\centering
\epsfig{file=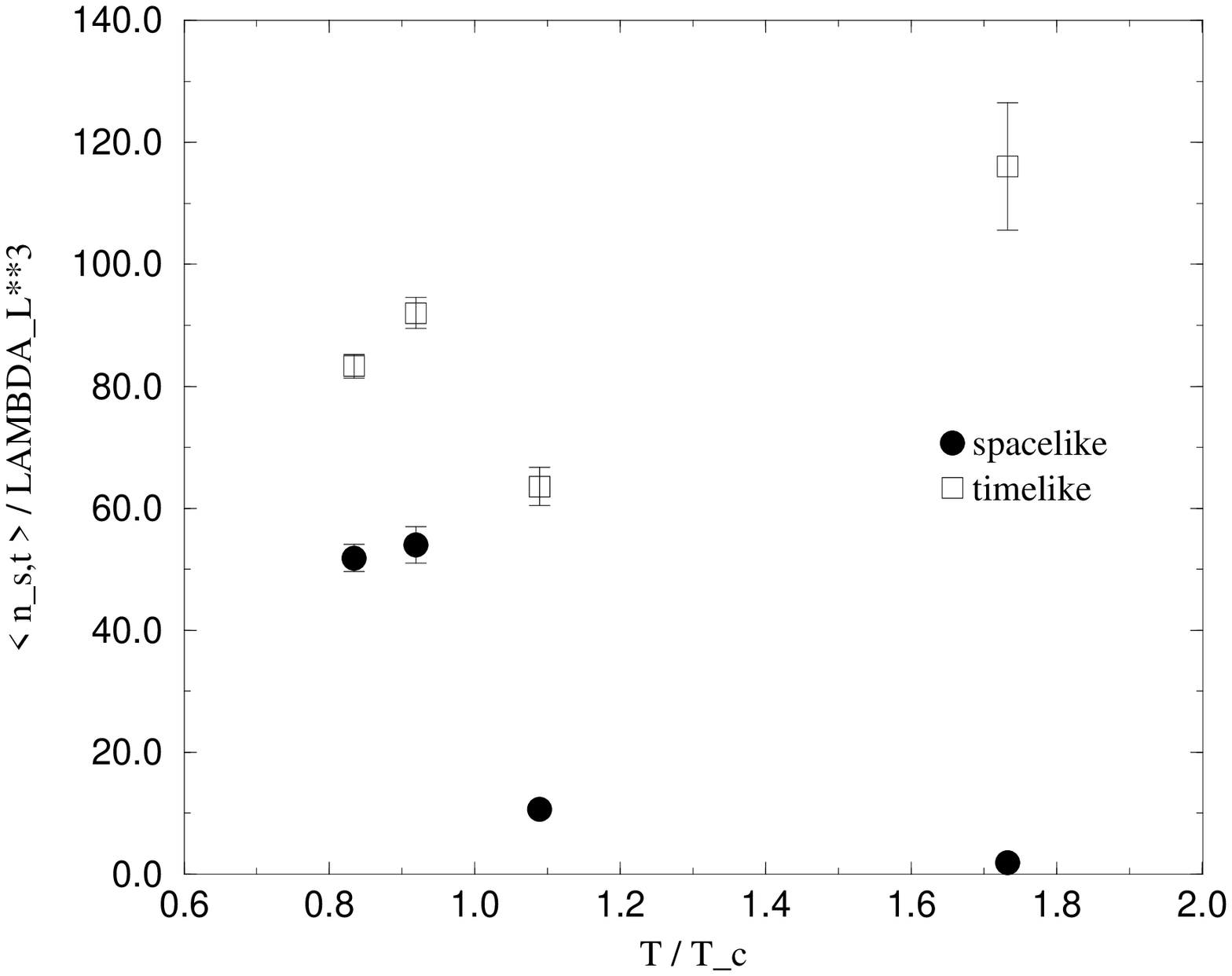,width=9.5cm,angle=0}
\caption{\sl $3$-dimensional densities of timelike and spacelike
monopoles across the
deconfining phase transition for inversely blocked configurations}
\label{fig:mondichten_ib}
\end{figure}

Fig. \ref{fig:mondichten_ib} expresses the monopole
currents for inversely blocked configurations
in the form of $3$-dimensional
densities
\begin{equation}
n_{\mu}= \frac{1}{N_{sites} a^3}~\langle M_{\mu} \rangle
\end{equation}
of timelike and spacelike
monopoles 
(in $\Lambda_L$ units)
as a function of $T/T_c$. Mo\-no\-po\-les, condensed in the confining phase,
become static (massive) particle--like objects
in the deconfined phase with a density
rising with temperature (cf. Ref.~\cite{Bornyakov2}~).
One may speculate about their relation to 't Hooft--Polyakov
monopoles.
Spacelike monopole currents (responsible for confinement below $T_{c}$) 
become strongly suppressed at $T_c$ and have
completely disappeared at $T \simeq 1.7~T_{c}$.

\section{Resolving the Vacuum Structure}

\subsection{Topological Density Correlation}

Information on the nature of topological excitations is contained
\begin{figure}[!thb]
\centering
\epsfig{file=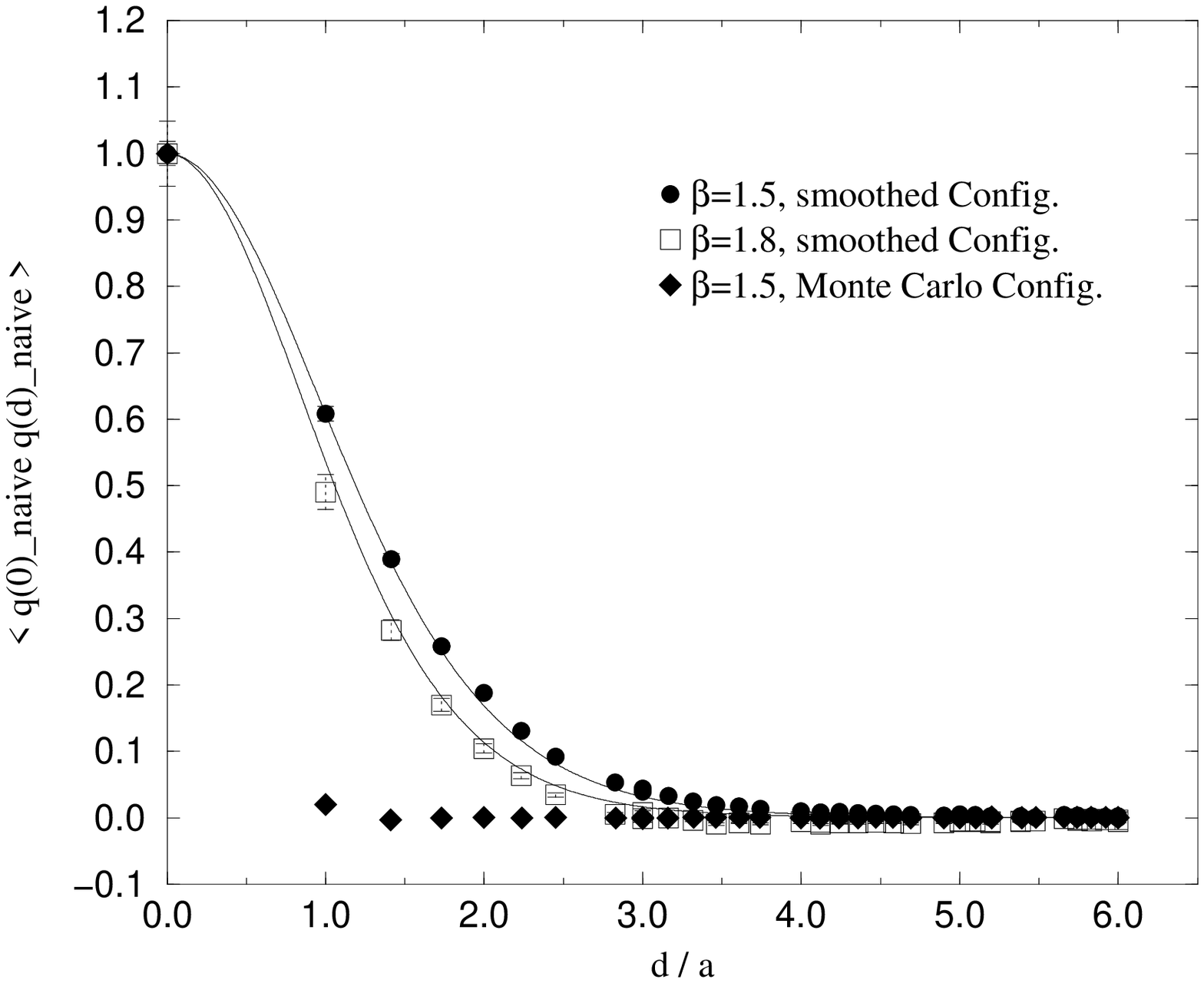,width=9.5cm,angle=0}
\caption{\sl Normalized correlation function of the naive
topological charge density at $\beta=1.5$ (for Monte Carlo and inversely 
blocked configurations)
and at $\beta=1.8$ (only inversely blocked). Solid lines are fits according
to eq. (22),(23)}
\label{fig:naive_corr}
\end{figure}
in the point--to--point correlation function of the topological charge
density. This is easy to measure, but for usual Monte Carlo configurations
$U^{MC}$ this measurement has no value.

Several groups have attempted to define shape and size of
topological excitations
by the use of cooling techniques. The MIT group \cite{Negele94}
has used for this
purpose deeply cooled configurations. 
There is an uncertainty about the loss of essential structures
during cooling. 
Sometimes there appears a particular structure just
during the first few cooling steps \cite{Vienna,IlgLATT94}. 
Recently, there
have been attempts to improve cooling by choosing 
suitably improved actions \cite{VanBaal,DeForcrand96,DeForcrand97}
for (unconstrained) relaxation. Partial success has been achieved to
stabilize instantons above some threshold in size. 
But certain
topological structures (instanton-antiinstanton pairs)
unavoidably disappear in the process of
cooling. 
The method
of inverse blocking
is welcome because it enables to detect correlations
on a set of well--defined configurations,
which are smooth but interpolate, in a locally correct way, coarser 
equilibrium
configurations. 

In Fig. \ref{fig:naive_corr} we compare the normalized
\begin{figure}[!thb]
\centering
\epsfig{file=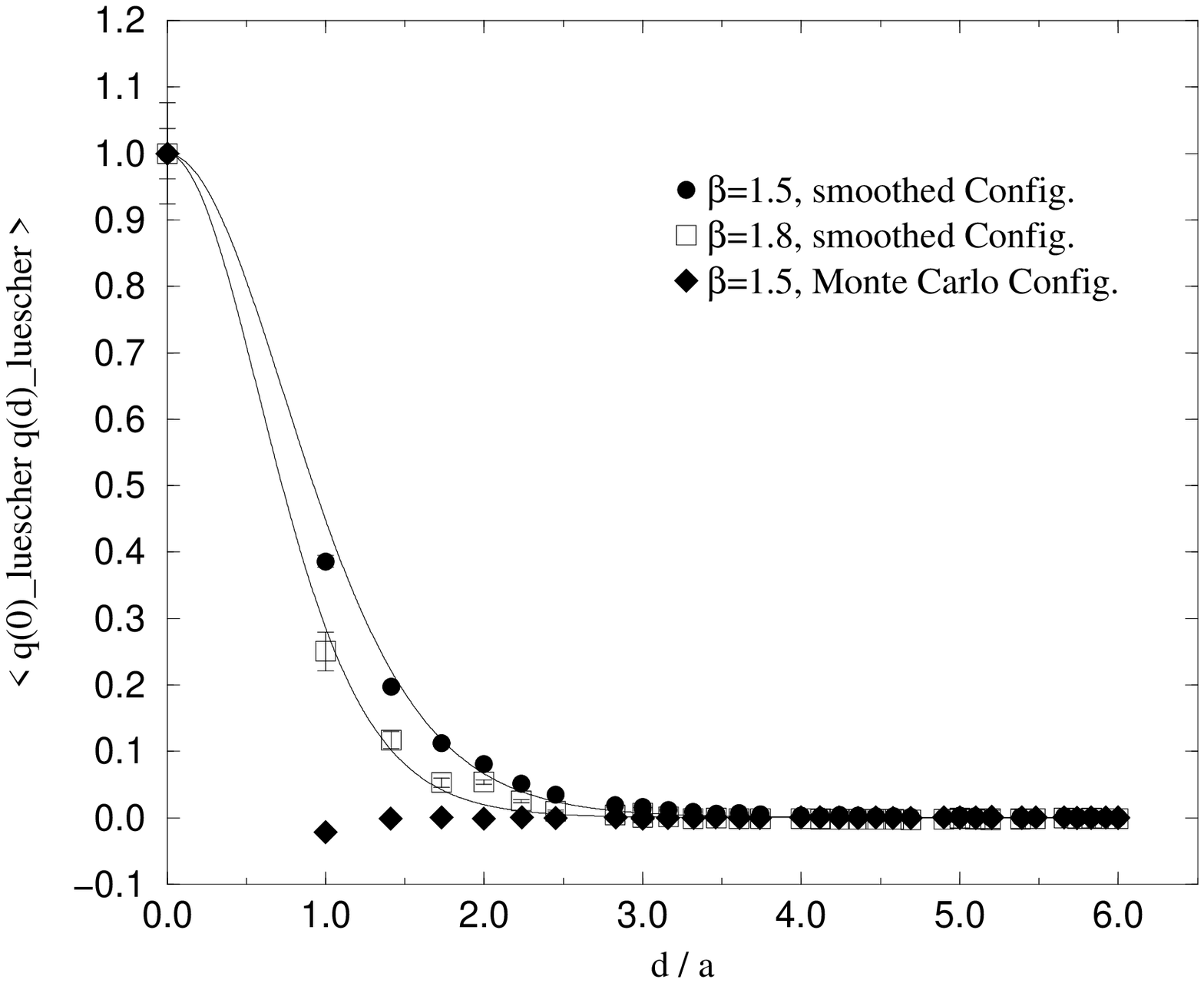,width=9.5cm,angle=0}
\caption{\sl Normalized correlation function of the geometric L\"uscher
topological charge density (as in Fig. 11)}
\label{fig:luescher_corr}
\end{figure}
correlation function of the topological charge
density measured on the $12^3\times 4$ lattice in the confinement phase
at $\beta=1.5$ on Monte Carlo and inversely blocked  
configurations. 
For normal Monte Carlo configurations 
there
is no signal
at any distance except for $x=0$. 
In order to demonstrate the influence of temperature on the signal
we show in this figure also the correlation function measured at
$\beta=1.8$ in the deconfinement phase. Due to the fixed
normalization at $x=0$ only
the change of the instanton radius can be seen (as explained below).

Smoothing is necessary in order to expose a signal in the
density--density correlation function not only in the case of
the naive 
topological density definition $q_{naive}(x)$.
Fig. \ref{fig:luescher_corr} shows the same for
L\"uscher's charge density. The effect is similar, but the correlation of
the naive charge density always extends somewhat farther. This can be
understood due to the extended nature of the operator (\ref{eq:q_naive}).

Getting a reasonable fit of
the correlation function to the analytical shape derived from the
instanton profile
\begin{equation}\label{eq:top_shape}
q_{inst}(x) = \frac{6}{\pi^2 \rho^4}
\left(\frac{\rho^2}{x^2 + \rho^2} \right)^4
\end{equation}
by folding
\begin{equation}\label{eq:folding}
\langle q(x) q(0) \rangle \propto \sum_{z} q_{inst}(x-z) q_{inst}(z)  ,
\end{equation}
we have determined the
average instanton radii $\rho_{inst}$ as given in Table
\ref{tab:radii}. The corresponding
fitting curves are shown in Figs. \ref{fig:naive_corr}
and \ref{fig:luescher_corr}. The instanton radius as determined by
this method decreases more rapidly with rising temperature 
than the inverse temperature.
Some change of $\rho_{inst}$ at deconfinement has been observed by
the cooling method, too \cite{ChuSchramm}.
\begin{table}[h]
\begin{center}
\begin{tabular}{|l|c|c|c|c|}
\hline
$\beta$                    & $1.50$ & $1.54$ & $1.61$ & $1.80$   \\
\hline
\hline
L\"uscher's charge   & $1.4~a$& $1.4~a$& $1.35~a$& $1.1~a$     \\
naive charge         & $1.8~a$& $1.8~a$& $1.75~a$& $1.6~a$     \\
\hline
\end{tabular}
\end{center}
\caption{\sl Instanton radii $\rho_{inst}$
in lattice units as estimated from the topological charge
density correlation function at two $\beta$--values in the confinement and
two $\beta$--values in the deconfinement phase,
according to the two charge definitions}
\label{tab:radii}
\end{table}

\subsection{Monopole--Instanton Correlation}

It has been demonstrated before for certain classical configurations
that there is a close connection between instantons and monopole world loops.
One can
visualize this in the maximally
Abelian gauge \cite{Bornyakov3}.
\begin{figure}[!thb]
\centering
\epsfig{file=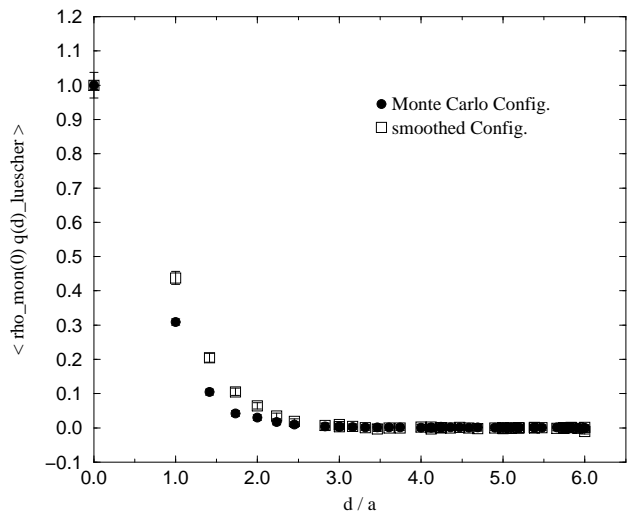,width=9.5cm,angle=0}
\caption{\sl Normalized correlation function of monopole
currents with topological charge density (according to L\"uscher)
at $\beta=1.5$ in the confinement phase}
\label{fig:qlumon_corr}
\end{figure}
A nontrivial correlation
has been measured between topological charge density and monopole
currents in the Monte Carlo ensemble of Euclidean field
configurations, too \cite{Vienna}.
We show here in Fig. \ref{fig:qlumon_corr}
the analogous normalized correlator for the original Monte Carlo
ensemble (generated with the fixed point action at $\beta=1.5$
in the confinement phase)
and for the corresponding ensemble of inversely blocked 
configurations. The latter correlator (for smoothed configurations)
is slightly wider.
In contrast to the topological density--density correlations, 
this correlator can be detected in the Monte Carlo configurations as well.

Comparing this with the same (normalized) correlation function
at $\beta=1.8$ in the deconfined phase (not shown here), 
we find that they are
identical as a function of the distance in lattice units
within the error bars. This means that this
correlation length changes proportional to the inverse
temperature.

\section{Conclusion and Outlook}

\begin{figure}[!thb]
\centering
\epsfig{file=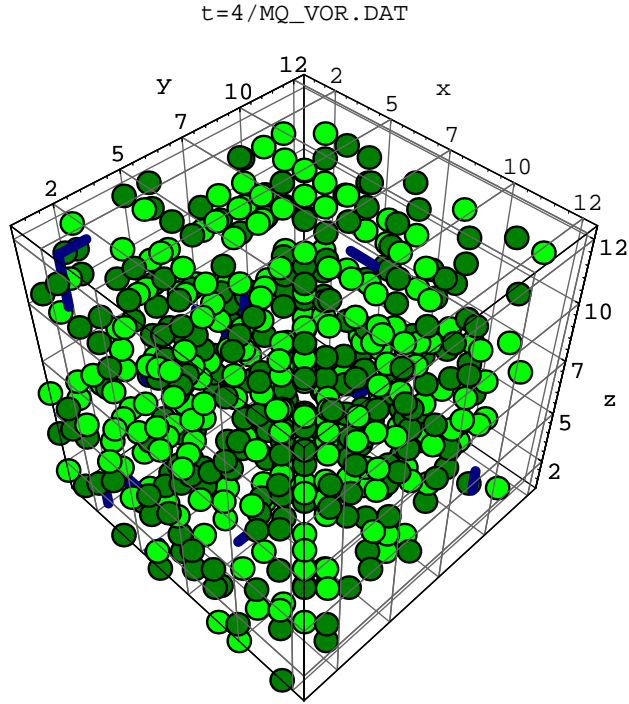,width=9.5cm,angle=0}
\caption{\sl
L\"uscher's topological charge density for $|q(x)|>q_0$
and monopole world lines in a timeslice of a
typical Monte Carlo configuration in the confinement phase;
signs of $q$ are distinguished by different color
}
\label{fig:unsmooth}
\end{figure}

We have explained in this paper how inverse blocking can be used
to analyze
topological
structures 
of genuine quantum configurations, 
usually hidden under ultraviolet fluctuations.

Apart from cooling techniques, inverse blocking is the only way
to measure autocorrelations of the topological density
giving information on the size of instantons. 
Information on monopole currents is available
also for normal Monte Carlo configurations.
Monopoles are correlated with the topological density,
both in Monte Carlo and inversely blocked configurations.
However, the density of monopole currents
is reduced, 
roughly proportional to the topological activity,
if a configuration undergoes the procedure
of
blocking followed by inverse blocking.

Only a small part of Abelian monopoles seen usually in
Monte Carlo configurations seems to be really important for
confinement. This part survives smoothing
in the confinement phase. We come to this conclusion
because the string tension is 
found to be largely unaffected
by the removal of ultraviolet fluctuations.
This observation supports
semiclassical scenarios of confinement.
In the deconfinement phase, however, we find that 
spacelike monopoles are strongly suppressed for inversely
blocked configurations while only timelike ones survive.

Armed with inverse blocking, we can do better than
measuring the topological density--density 
correlation
or the correlation between monopoles and topological density.
We are now able to study 
clustering properties of the topological charge using
inversely blocked lattice configurations.
We can relate
clusters of topological charge to the 
diluted network of monopole currents.
Instead of examining idealized semiclassical configurations
under different aspects of topological and monopole structure,
the method of inverse blocking makes it possible to analyze generic
Euclidean
lattice field histories in order to characterize
the physical phase one is dealing with.
The results should be interesting for vacuum model builders.
In view of chiral symmetry breaking and restoration,
it would be also interesting to study the influence of the inverse
blocking RG
transformation on the
spectrum of the Dirac operator.

\begin{figure}[!thb]
\centering
\epsfig{file=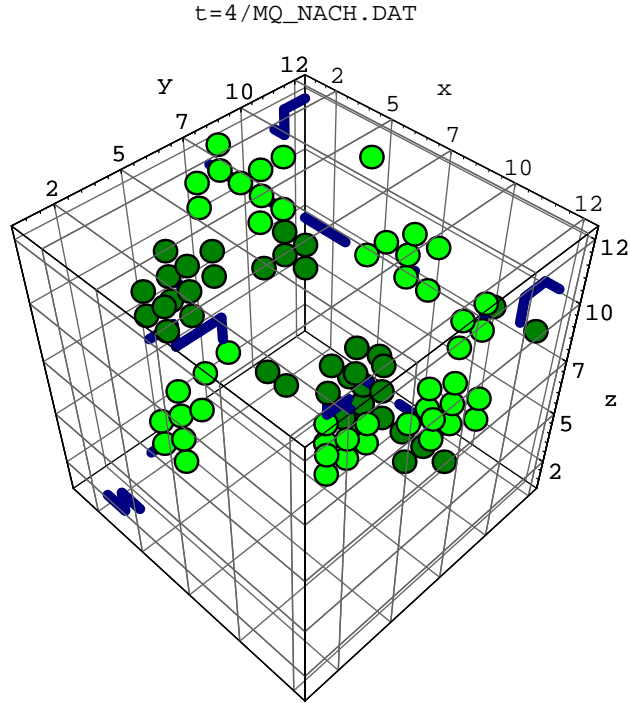,width=9.5cm,angle=0}
\caption{\sl
Clustering of the L\"uscher topological charge density
and monopole world lines in a timeslice of
the same configuration as in Fig. \ref{fig:unsmooth} but after smoothing
}
\label{fig:smooth_luescher}
\end{figure}

The study of clusters of topological charge promises to 
give a more direct access to the parameters and correlations
that should be built in an instanton model for QCD at zero and finite
temperature.
In Figs. \ref{fig:unsmooth}
and \ref{fig:smooth_luescher}
lattice hypercubes
of one timeslice are marked  
if the L\"uscher topological charge density exceeds a threshold
value $q_0=0.015$. Different colors symbolize the sign.
The original Monte Carlo configuration does not allow to guess any structure.
The other picture shows the structure emerging after blocking and inverse
blocking.
The effect is the same 
using either naive or L\"uscher's charge density. Therefore one can
abandon to use complicated algorithms 
and use naive or improved loop--oriented 
definitions of topological charge density.
\par
\medskip
\noindent
There are two immediate directions for further studies.
\begin{itemize}
\item{} Study of the clustering properties of topological
charge in individual lattice configurations 
and of the relation to the monopole currents. 
\item{} Repeating the present explorative study at higher $\beta$
with a particular emphasis on
the scale dependence of topological structure.
\end{itemize}
Further progress in the construction of the perfect action should be
incorporated into this program.

\section*{Note added:}

During the time elapsed since the first submission of our article,
several interesting papers have appeared which have attempted to
investigate 
the space time topological structure of lattice gauge fields. Three
methods have been used: (improved) cooling, fermionic
(spectral) methods and smoothing. The latter method had been proposed 
in this paper
as a general, renormalization group oriented pattern recognition method
for large scale structures in terms of topological charge and monopoles
which conserves the string tension. 

While the subsequent studies were mostly focussed on 
vacuum (zero temperature) configurations,
the results are still controversial as emphasized in Ref. \cite{Narayanan1}. 
As a generic rule, one can state that
improved cooling \cite{DeForcrand97} is biased towards larger instantons
(because only those are stable)
while the topological susceptibility tends to be lower; in contrast, the
smoothing method (as developed further by the Boulder group \cite{DeGrand97})
favours smaller instanton sizes and tends to find somewhat higher
topological susceptibility. This applies to $SU(2)$ pure gauge theory
where the results can be compared (but not without additional assumptions).

We want to stress an important difference of Ref. \cite{DeGrand97} 
to the work published here. 
Smoothing has been applied there in an iterative way 
that leads finally
to locally classical configurations. The number
of iterations appears there as somewhat subjective parameter similar to
the number of (unimproved or improved) cooling steps.

This goes far beyond our intention. From our point of view, 
the advantage of smoothing is
that it provides a minimal smoothing  
of quantum fluctuations
in clear correspondence to a certain scale. 
This scale is well--defined by  
the highest blocking level.   

The effect of smoothing on various Abelian projections has been critically
examined in Ref. \cite{Kovacs}. The maximally Abelian gauge is the only one
which seems to exhibit long range physics also if applied to smoothed 
lattice fields.
While we had concentrated on the density of monopoles, it has been found there
that the property of Abelian dominance ({\it i.e.}
dominance of the Abelian string tension) is only
partly preserved.

A large number of investigations was studying topological structures 
using fermionic methods going back to Smit and Vink \cite{SmitVink}. 
It is important in the context of this work that these methods seem 
to work immediately, 
without smoothing or cooling techniques which otherwise would have to be 
applied to the Monte Carlo lattice gauge field configurations. 
Investigating the spectral
flow of the lowest eigenvalues of the Dirac operator
\cite{Narayanan1,Narayanan2,Simma} these
authors are able to
count the number of localized lumps (and identify the respective
sign) of topological charge. 
The local topological structure is thereby indirectly
accessible through the localization properties of the corresponding 
eigenmodes. This has been checked in Refs. \cite{Simma,Negele97}
with cooling, but there is no comparison so far with the topological
structure detected by smoothing. 

We have recently improved our relaxation algorithm for implementing inverse
blocking \cite{InProgress}. 
We are now in the position to optimize a fixed point action of moderate
complexity
for a given inverse blocking parameter $\kappa$. We could check a 
new parametrization 
of the perfect action for $SU(2)$ which is truncated compared with the action
used in Ref. \cite{DeGrand97}. We thank the Boulder group for correspondence
about that and for providing the truncated action \cite{PrivateCommunication}.
We can now show that the parametrization of Table \ref{tab:weights} is not
compatible with the $\kappa=12$ required. This inconsistency had forced
us to accept (in this paper) a smaller effective $\kappa$. We have 
convinced ourselves that the statistical observations are 
independent of this choice although the inversely blocked configurations
may differ locally.


\begin{thebibliography}{99}

\bibitem{CDG}
C. G. Callan, R. Dashen and D. J. Gross, 
Phys.~Rev.~\bf D 17, \rm (1978) 2717;
D. J. Gross, R. D. Pisarski and L. G. Yaffe,
\sl Rev.~Mod.~Phys. \bf 53, \rm (1981) 43 
\bibitem{Shuryak}
E.~Shuryak, 
Phys.~Reports~\bf 264, \rm (1996) 357;
T. Sch\"afer and E. V. Shuryak
Phys.~Rev.~\bf D 54, \rm (1996) 1099; 
T. Sch\"afer and E. V. Shuryak,
Phys.~Rev.~\bf D 53, \rm (1996) 6522; 
T. Sch\"afer and E. V. Shuryak,
\sl e-print archive \rm hep-ph/9610451
\bibitem{Mandelstam}
S. Mandelstam,
\sl Phys.~Reports~\bf 23C, \rm (1976) 245 
\bibitem{tHooft}
G. 't Hooft,
Nucl.~Phys.~\bf B 190, \rm (1981) 455 
\bibitem{Improved}
F. Niedermayer,
\sl Nucl.~Phys.~\bf B \sl Proc. Suppl.~\bf 53 \rm (1997) 56;
W. Bietenholz, R. Brower, S. Chandrasekharan and U. J. Wiese,
\sl Nucl.~Phys.~\bf B \sl Proc. Suppl.~\bf 53 \rm (1997) 921
\bibitem{FixedPointG}
T. A. DeGrand, A. Hasenfratz, P. Hasenfratz, F. Niedermayer,
and U. J. Wiese, 
\sl Nucl.~Phys.~\bf B \sl Proc. Suppl.~\bf 42 \rm (1995) 67; 
T. DeGrand, A. Hasenfratz, P. Hasenfratz and F. Niedermayer,
\sl Nucl.~Phys.~\bf B 454, \rm (1995) 587;
\sl Nucl.~Phys.~\bf B 454, \rm (1995) 615; 
\sl Phys.~Lett.~\bf B 365, \rm (1996) 233; 
M. Blatter and F. Niedermayer,
\sl Nucl.~Phys.~\bf B 482, \rm (1996) 286 
\bibitem{FixedPointF}
T. DeGrand, A. Hasenfratz, P. Hasenfratz, P. Kunszt and F. Niedermayer,
\sl Nucl.~Phys.~\bf B \sl Proc. Suppl.~\bf 53 \rm (1997) 942 
\bibitem{DeGrand96.1}
T. DeGrand, A. Hasenfratz and De-cai Zhu,
\sl Nucl.~Phys.~\bf B 475, \rm (1996) 321 
\bibitem{DeGrand96.2}
T. DeGrand, A. Hasenfratz and De-cai Zhu,
\sl Nucl.~Phys.~\bf B 478, \rm (1996) 349 
\bibitem{Cooling}
E.--M. Ilgenfritz, M. L. Laursen, M. M\"uller--Preussker,
G. Schierholz and H. Schiller,
\sl Nucl.~Phys.~\bf B 168, \rm (1986) 693; 
M. Teper,
\sl Phys.~Lett.~\bf B 162, \rm (1985) 357
\bibitem{Bornyakov1}
V. G. Bornyakov, E.--M. Ilgenfritz, M. L. Laursen,
V. K. Mitrjushkin, M. M\"uller--Preussker and A. J. van der Sijs,
\sl Phys.~Lett.~\bf B 261 \rm (1991) 116
\bibitem{Bornyakov2}
V. G. Bornyakov, V. K. Mitrjushkin and
M. M\"uller-Preussker,
\sl Phys. Lett.~\bf B 284, \rm (1992) 99
\bibitem{Kronfeld}
A. S. Kronfeld, G. Schierholz and U. J. Wiese,
\sl Nucl.~Phys.~\bf B 293, \rm (1987) 461; 
A. S. Kronfeld, M. L. Laursen, G. Schierholz and
U. J. Wiese,
\sl Phys.~Lett.~\bf B 198, \rm (1987) 516
\bibitem{DeGrandToussaint}
T. A. DeGrand and D. Toussaint,
\sl Phys.~Rev.~\bf D 22, \rm (1980) 2478 
\bibitem{Bornyakov3}
V. G. Bornyakov and G. Schierholz,
\sl Phys.~Lett.~\bf B 384, \rm (1996) 190; 
M. N. Chernodub and F. V. Gubarev,
\sl JETP Lett.~\bf 62, \rm (1995) 100; 
A. Hart and M. Teper,
\sl Phys.~Lett.~\bf B 371, \rm (1996) 261; 
R. C. Brower, K. N. Orginos and C.--I Tan,
\sl Nucl.~Phys.~\bf B \sl Proc. Suppl.~\bf 53 \rm (1997) 488 and 
\sl e-print archive   
hep-th/9610101
\bibitem{Vienna}
S. Thurner, M. Feurstein, H. Markum and W. Sakuler,
\sl Phys.~Rev.~\bf D 54, \rm (1996) 3457; 
S. Thurner, H. Markum and W. Sakuler,
in {\it Proceedings of Confinement 95}, Osaka 1995, eds. H.~Toki et al.
(World Scientific, 1996) 77; 
H. Markum, W. Sakuler and S. Thurner,
\sl Nucl.~Phys.~\bf B \sl Proc. Suppl.~\bf 47 \rm (1996) 254 
\bibitem{PhillipsStone}
A. Phillips and D. Stone,
\sl Comm.~Math.~Phys.~\bf 103, \rm (1986) 599; 
A. S. Kronfeld, M. L. Laursen, G. Schierholz,
C. Schleiermacher and U. J. Wiese,
\sl Comp.~Phys.~Comm.~\bf 54, \rm (1989) 109
\bibitem{Luescher}
M. L\"uscher,
\sl Comm.~Math.~Phys.~\bf 85, \rm (1982) 29
\bibitem{LuescherImplem}
I. A. Fox, J. P. Gilchrist, M. L. Laursen
and G. Schierholz,
\sl Phys. Rev.~Lett.~\bf 54, \rm (1985) 749; 
A. S. Kronfeld, M. L. Laursen, G. Schierholz and U. J. Wiese,
\sl Nucl.~Phys.~\bf B 292, \rm (1987) 330
\bibitem{DiVecchia}
P. Di Vecchia, K. Fabricius, G. C. Rossi and G. Veneziano,
\sl Nucl.~Phys.~\bf B 192, \rm (1981) 392;
\sl Phys.~Lett.~\bf B 108, \rm (1982) 323;
\sl Phys.~Lett.~\bf B 249, \rm (1990) 490
\bibitem{MMPMakh}
N. V. Makhaldiani and M. M\"uller--Preussker,
\sl JETF Pisma~\bf 37, \rm (1983) 440
\bibitem{NoScaling}
M. G\"ockeler, A. S. Kronfeld, M. L. Laursen,
G. Schierholz and U. J. Wiese,
\sl Phys.~Lett.~\bf B 233, \rm (1989) 192
\bibitem{DiGiacomoT1}
A. Di Giacomo, E. Meggiolaro
and H. Panagopoulos,
\sl Phys.~Lett.~\bf B 277, \rm (1992) 491
\bibitem{IlgLATT94}
E.--M. Ilgenfritz, E. Meggiolaro and
M. M\"uller--Preussker,
\sl Nucl.~Phys.~\bf B \sl Proc. Suppl.~\bf 42 \rm (1995) 496 
\bibitem{ImprovedQ}
C. Christou, A. Di Giacomo, H. Panagopoulos
and E. Vicari,
\sl Phys.~Rev.~\bf D 53, \rm (1996) 2619
\bibitem{DiGiacomoT2}
B. Alles, M. D'Elia and A. Di Giacomo,
\sl Nucl.~Phys.~\bf B \sl Proc. Suppl.~\bf 53 \rm (1997) 541
and
\sl e-print archive \rm 9605013
\bibitem{HasenfratzNPB454I}
see the second of Ref. \cite{FixedPointG} 
\bibitem{Smearing}
M. Falcioni, M. Paciello, G. Parisi and B. Taglienti,
\sl Nucl.~Phys.~\bf B 251, \rm (1985) 624; 
M. Albanese et al.,
\sl Phys.~Lett.~\bf B 192, \rm (1987) 163
\bibitem{DiGiacomoString}
M. Campostrini, A. Di Giacomo, M. Maggiore
H. Panagopoulos and
E. Vicari,
\sl Phys.~Lett.~\bf B 225, \rm (1989) 403
\bibitem{DiGiacomoTopcool}
M. Campostrini, A. Di Giacomo and H. Panagopoulos,
\sl Phys.~Lett.~\bf B 212, \rm (1988) 206; 
M. Campostrini, A. Di Giacomo, H. Panagopoulos and E. Vicari,
\sl Nucl.~Phys.~\bf B 329, \rm (1990) 683
\bibitem{Negele94}
M. C. Chu, J. M. Grandy, S. Huang and J. W. Negele,
\sl Phys.~Rev.~\bf D 49, \rm (1994) 6039; 
R. C. Brower, T. L. Ivanenko, J. W. Negele and K. N. Orginos,
\sl Nucl.~Phys.~\bf B \sl Proc. Suppl.~\bf 53 \rm (1997) 547 
\bibitem{Renormalization}
A. Di Giacomo and E. Vicari,
\sl Phys.~Lett.~\bf B 275, \rm (1992) 429; 
B. Alles, M. Campostrini, A. Di Giacomo,
Y. G\"und\"uc and E. Vicari,
\sl Phys.~Rev.~\bf D 48, \rm (1993) 2284
\bibitem{VanBaal}
M. Garcia Perez, A. Gonzalez--Arroyo, J. Snippe
and P. van Baal,
\sl Nucl.~Phys.~\bf B 413 \rm (1994) 535
\bibitem{DeForcrand96}
P. de Forcrand, M. Garcia Perez and I.--O. Stamatescu,
\sl Nucl.~Phys.~\bf B \sl Proc. Suppl.~\bf 47 \rm (1996) 278 and 777
\bibitem{DeForcrand97}
P. de Forcrand, M. Garcia Perez and I.--O. Stamatescu,
\sl Nucl.~Phys.~\bf B 499 \rm (1997) 409;
P. de Forcrand, M. Garcia Perez, J. E. Hetrick and I.--O. Stamatescu,
\sl contribution to Lattice 97,
e-print archive \rm  hep-lat/9710001
\bibitem{1/N}
E.--M. Ilgenfritz and M. M\"uller--Preussker,
\sl Phys.~Lett.~\bf B 99, \rm (1981) 128
\bibitem{Muenster}
G. M\"unster,
\sl Z.~f.~Phys.~\bf C 12, \rm (1982) 43
\bibitem{MonopolCondensation}
L. Del Debbio, A. Di Giacomo,
G. Paffuti and P. Pieri,
\sl Nucl.~Phys.~\bf B \sl Proc. Suppl.~\bf 42 \rm (1995) 234 and 
\sl Phys.~Lett.~\bf B 355, \rm (1995) 255; 
M. N. Chernodub, M. I. Polikarpov and A. I. Veselov,
\sl Nucl.~Phys.~\bf B \sl Proc. Suppl.~\bf 49 \rm (1996) 307 and 
\sl e-print archive \rm hep-lat/9610007;
A. I. Veselov, M. I. Polikarpov and M. N. Chernodub,
\sl JETP Lett. \bf 63, \rm (1996) 411 
\bibitem{ChuSchramm}
M. C. Chu and S. Schramm,
\sl Phys.~Rev.~\bf D 51, \rm (1995) 4580
\bibitem{DeGrand97}
T. DeGrand, A. Hasenfratz, T. G. Kovacs,
\sl e-print archive \rm hep-lat/9705009; and 
\sl contribution to Lattice 97,
e-print archive \rm hep-lat/9709095 
\bibitem{Kovacs}
T. G. Kovacs, Z. Schram,
\sl e-print archive \rm hep-lat/9706012, 
\sl Phys.~Rev.~\bf D \rm (to be published) 
\bibitem{SmitVink}
J. Smit and J. Vink, 
\sl Nucl.~Phys.~\bf B 286, \rm (1987) 485, \bf B 303, \rm (1988) 36
\bibitem{Narayanan1}
R. Narayanan and R. L. Singleton,
\sl contribution to Lattice 97,
e-print archive \rm  hep-lat/9709014 
\bibitem{Narayanan2}
R. Narayanan and P. Vranas,
\sl e-print archive \rm hep-lat/9702005, \sl Nucl.~Phys.~\bf B \rm 
(to be published)
\bibitem{Simma}
D. Smith, H. Simma and M. Teper (UKQCD Collaboration),
\sl contribution to Lattice 97,
e-print archive \rm  hep-lat/9709128 
\bibitem{Negele97}
T. L. Ivanenko and J. W. Negele, 
\sl contribution to Lattice 97,
e-print archive \rm  hep-lat/9709130; 
J. W. Negele, 
\sl contribution to Lattice 97,
e-print archive \rm  hep-lat/9709129 
\bibitem{InProgress}
E.--M.~Ilgenfritz,
M.~M\"uller--Preussker
and S.~Thurner, \sl in progress \rm
\bibitem{PrivateCommunication}
T. G. Kovacs, \sl private communication \rm



\end{thebibliography}
\end{document}